\documentclass[fleqn,10pt]{wlscirep}
\usepackage[utf8]{inputenc}
\usepackage[T1]{fontenc}
\usepackage{siunitx}
\usepackage{amssymb}
\usepackage{hyperref}
\usepackage{float}
\usepackage{wasysym}

\title{\textit{In situ} theranostic platform uniting highly localized magnetic fluid hyperthermia, magnetic particle imaging, and thermometry in 3D}
 
\author[1,2]{Oliver Buchholz}
\author[3]{Kulthisa Sajjamark}
\author[3]{Jochen Franke}
\author[4]{Huimin Wei}
\author[4]{André Behrends}
\author[1,2]{Christian Münkel}
\author[5]{Cordula Grüttner} 
\author[6]{Pierre Levan} 
\author[2,7]{Dominik von Elverfeldt} 
\author[4,8]{Matthias Gräser}
\author[4,8]{Thorsten Buzug} 
\author[1,2,7,+*]{Sébastien Bär}
\author[1,2,+]{Ulrich G. Hofmann}

\affil[1]{Section for Neuroelectronic Systems, Department of Neurosurgery, University Medical Center Freiburg, Freiburg, Germany}
\affil[2]{Faculty of Medicine, University of Freiburg, Freiburg, Germany}
\affil[3]{Bruker BioSpin MRI GmbH, Preclinical Imaging Division, Ettlingen, Germany}
\affil[4]{Fraunhofer Research Institution for Individualized and Cell-Based Medical Engineering IMTE, Lübeck, Germany}
\affil[5]{micromod Partikeltechnologie GmbH, Rostock, Germany}
\affil[6]{Department of Radiology and Hotchkiss Brain Institute, Cumming School of Medicine, University of Calgary, Calgary, AB, Canada}
\affil[7]{Division of Medical Physics, Department of Diagnostic and Interventional Radiology, University Medical Center Freiburg, Freiburg, Germany}
\affil[8]{Institute of Medical Engineering, University of Lübeck, Germany}

\affil[*]{sebastien.baer@uniklinik-freiburg.de}
\affil[+]{these authors contributed equally to this work}

\begin{document}
\begin{abstract}

In all of medical profession a broad field of research is dedicated to seek less invasive and low-risk forms of therapy with the ultimate goal of non-invasive therapy, particularly in neoplasmic diseases. Theranostic platforms, combining diagnostic and therapeutic approaches within one system, have thus garnered interest to augment invasive surgical, chemical, and ionizing interventions. Magnetic particle imaging (MPI) offers, with its versatile tracer material (superparamagnetic iron oxide nanoparticles, SPION), a quite recent alternative to established radiation based diagnostic modalities. In addition, MPI lends a bimodal theranostic frame allowing to combine tomographic imaging with therapy techniques using the very same SPION. In this work, we show for the first time the interleaved combination of MPI-based imaging, therapy (highly localized magnetic fluid hyperthermia) and therapy safety control (MPI-based thermometry) within one theranostic platform in all three spatial dimensions.

\end{abstract}
\flushbottom
\maketitle
\thispagestyle{empty}
\section*{Introduction}

Diagnosis and therapy are the two sides of the same medal applied by the medical professions. Usually, they are executed asynchronously, but with an improvement of technology it became feasible to put the coin back into one palm and the neologism \textit{"theranostics"} was created \cite{Simon2021,Wiesing2019,Gilham2002}. Even though it originally aimed to explain dual use of substances e.g. in nuclear medicine, point of care diagnostic and drug delivery \cite{Gomes2020,Gilham2002}, \textit{theranostics} does propose instrumentation capable of both imaging and treatment in one device \cite{Janib2010} which we report on in the following enabled by a quite recent imaging modality: Magnetic Particle Imaging integrated with Magnetic Fluid Hyperthermia. 
Magnetic particle imaging (MPI) is a tomographic imaging technique allowing for non-invasive quantification of versatile single domain iron oxide nanoparticles (SPION) \cite{Herynek2021, Yang2022}. The foundational concept of MPI was introduced in 2005 entailing an alternating magnetic field (drive field), which allows to change the magnetization of SPION\cite{Gleich}. Hereby, the intrinsic superparamagnetic properties of these SPIONs are exploited, as their non-linear magnetization response results in a distorted signal which can be detected by a receive system. The induced SPION signal, analyzed in the frequency domain, contains a plurality of harmonics of the drive frequency \cite{Rahmer:2009uh}. This induced signal scales linearly with the SPION concentration \cite{BORGERT2012} and thus makes MPI a quantitative method. By superimposing a static magnetic gradient field (selection field) vanishing at the center (field free region, FFR), the SPION signal can be encoded in space to permit for tomographic imaging  \cite{Gleich}. This is a consequence of the SPIONs' magnetization curve as they magnetically saturate outside the FFR not contributing to the received signal \cite{Gleich}. In contrast, magnetic SPIONs inside the FFR contribute distinctively to the received signal and thus permit tomographic image formation \cite{Gleich}. During image acquisition, the FFR is moved in space by two additional superimposed magnetic fields in a predefined trajectory spanning up an accessible field-of-view (FOV)\cite{Rahmer:2009uh, Wells2020}. 
Since only the magnetic SPIONs' specific magnetization properties contribute to the image signals, MPI displays high sensitivity towards these magnetic SPIONs and a background free contrast  \cite{Zhou2018}. \textit{In vivo} SPIONs remain superparamagnetic until hydrolyzed, enzymatically degraded \cite{Keselman_2017} or excreted along the mononuclear phagocytic system \cite{Arami2015,Nowak2022} offering the possibility for long term monitoring depending on their metabolic properties  \cite{Liu2021}. Also, changes in the SPION's environment such as its temperature, viscosity, and pH \cite{Rauwerdink2010,Rauwerdink2009,Rauwerdink2010oct} influence their magnetization. This opens the door for a wide range of applications \cite{BuchFranke} including MPI-based thermometry \cite{Stehning, Buchholz2022}. 
While contributing to image generation in the FFR, magnetic SPIONs may become subject to further physical manipulations. For example, high frequency alternating magnetic fields (AMF) may lead to energy dissipation (heating) caused by magnetization losses governed by internal Néel fluctuations of (single domain) SPION magnetic moments, external Brownian fluctuations and potentially magnetisation hysteresis \cite{MA200433}. This heating procedure by alternating magnetic fields (AMF) is called magnetic fluid hyperthermia (MFH) \cite{ROSENSWEIG2002}. By exploiting the same spatial encoding mechanism as in MPI, the volume in which magnetic SPIONs contribute to MFH can be confined in 3D by the same magnetic gradient field to the vicinity of the FFR \cite{Wang2005}. Heating efficiency is determined by SPION's specific-absorption rate (SAR) and is dependent on shape anisotropy \cite{guardia}, coating material\cite{Jordan} and thickness\cite{Liu}, particle size\cite{Tong}, and thermal conductivity of the surface coating and solvent viscosity\cite{Prashant}. From an instrumentation point of view, the SAR is governed by the amplitude and frequency of the AMF excitation while the spatial specificity is dominated by the magnetic field gradient slope.  
Wedding magnetic particle imaging and magnetic fluid hyperthermia could thus offer unprecedented benefits for the field of \textit{theranostics} and was indeed already attempted in back to back instruments to selectively target liver tumors \textit{in vivo} without substantially heating surrounding tissue \cite{Tay}.  
In the study at hand we go beyond this success and demonstrate tomographic MPI interleaved with locally restricted MFH  (MPI-MFH)  while controlling for achieved temperatures with MPI-based thermometry within the same platform and session.

\begin{figure}[ht!]
\centering
\includegraphics[width=\linewidth]{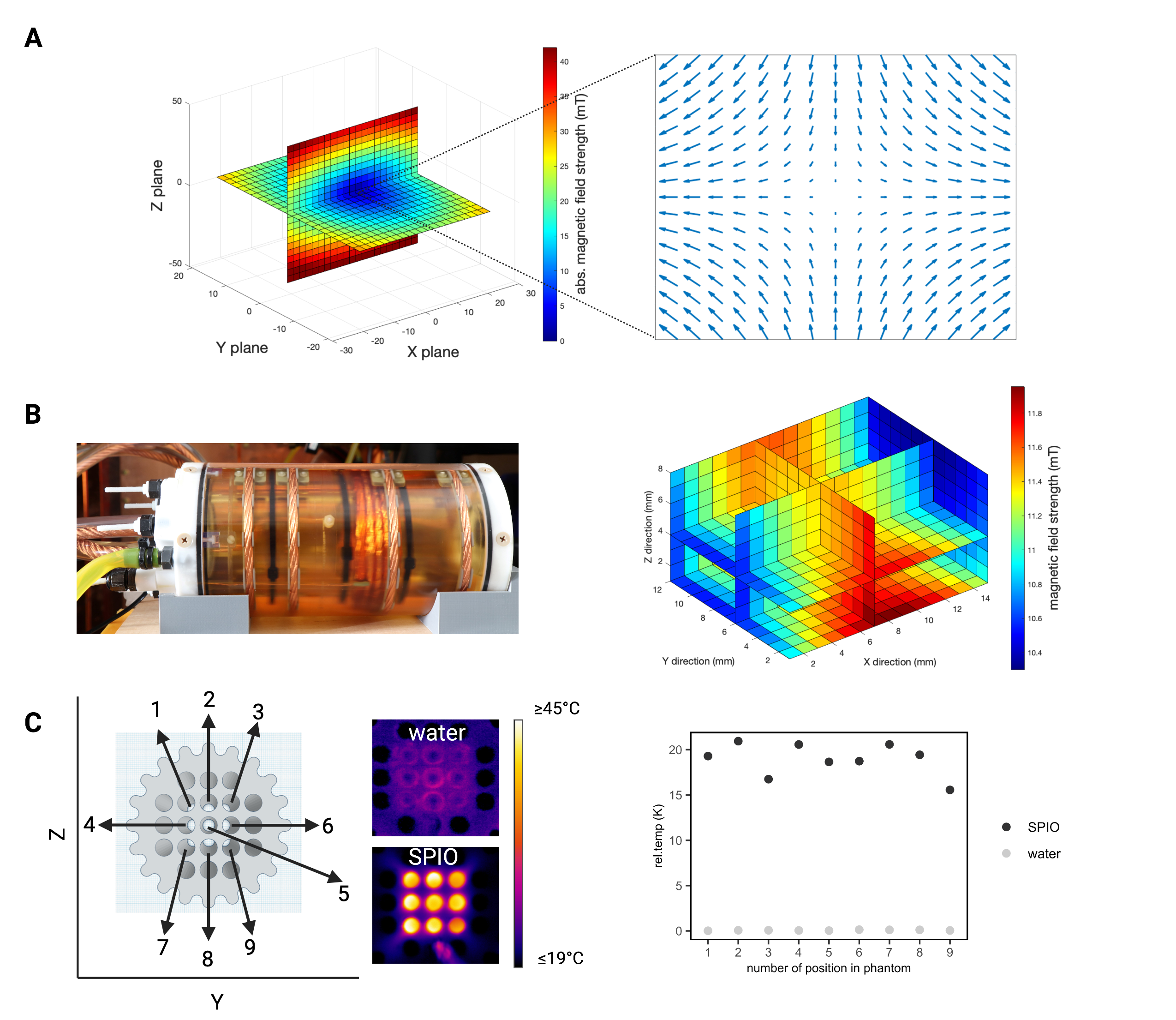}
\caption{\textbf{MPI guided MFH} (A) Visualization of the gradient field distribution in space resulting from spatially varying magnetic focus fields. Only SPION located in the region with the vanishing magnetic strength (field free region, FFR) can interact (imaging, heating, etc.) with additional magnetic influences.  (B) Calculated magnetic field strength at the center of the hyperthermia insert over a volume of 22.5 mm x 18 mm x 12 mm (x,y,z) in steps of 1.5 mm in each direction (C). FLIR-camera characterization of global heating (focus fields OFF)  and spatial distribution in a 3D phantom of SPION samples within the FOT distributed over a grid of 16x16mm (in y-and z- direction) surrounding the center of the MPI-MFH platform (x =0)}
\label{fig:MPI guided MFH}
\end{figure}

\section*{Results}
\subsection*{General description of our MPI-based theranostic system}

\begin{figure}[ht!]
\centering
\includegraphics[width=\linewidth]{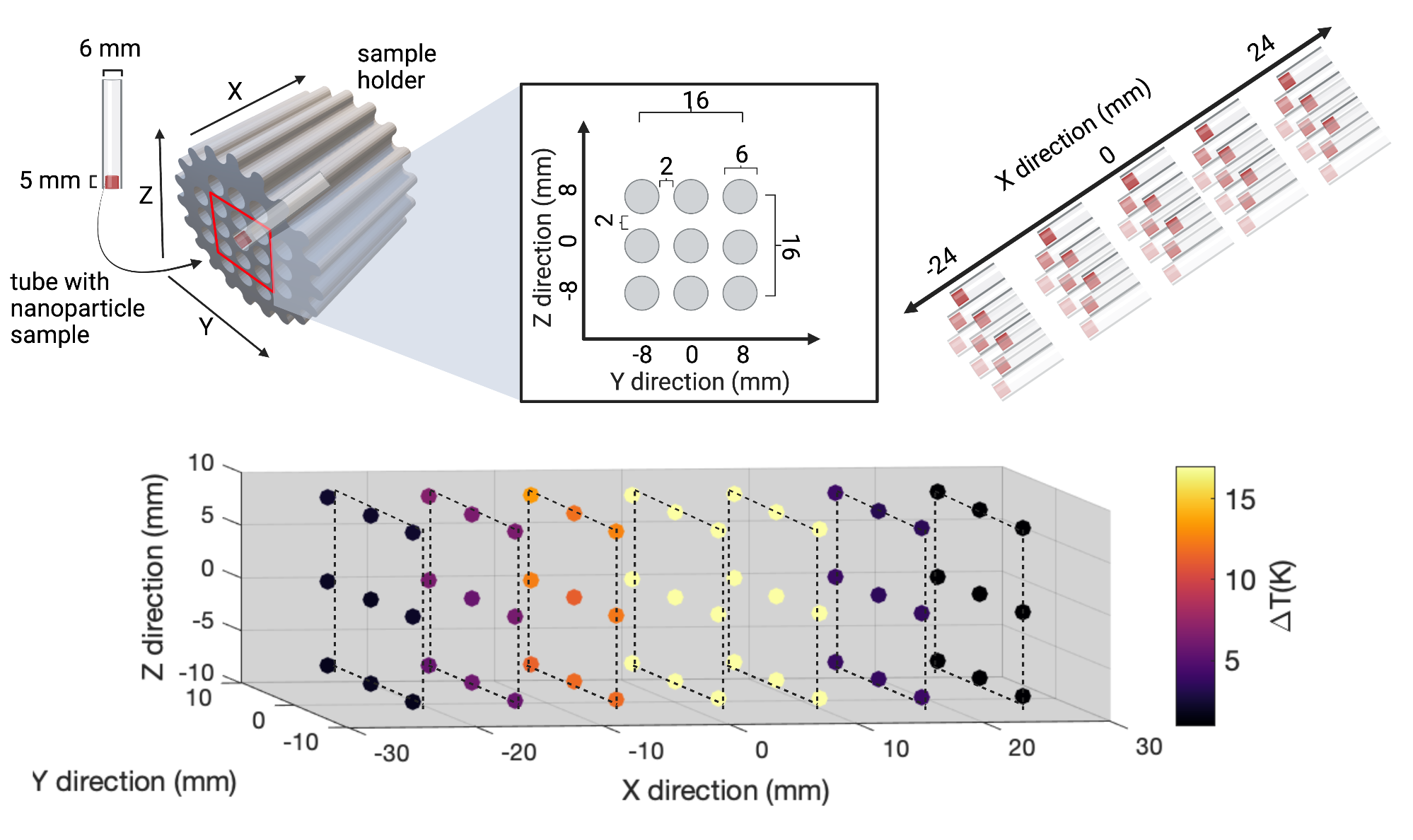}
\caption{\textbf{Characterization of field of therapy} The heating efficiency of SPION at distinct spatial locations within the MPI-MFH platform was assessed. For that, 9 thin walled glass tubes, each filled with \SI{50}{\micro\liter} SPION (synomag-S-90) and topped with paraffin were placed horizontally inside a 3D-printed sample holder. The bottom of the samples was facing towards a thermal camera in approximately \SI{1.5}{m} distance. The 9 samples were distributed equidistantly across the cross section at the relative center of the hyperthermia insert. Subsequently, AMF was applied (focus field gradients OFF, i.e. global MFH subjected to the entire FOT). By moving the sample holder in defined increments of \SI{8}{mm} along the x plane (both in -x and x direction), the extent of heating in x, y, z direction was determined.}
\label{fig:globalMFH}
\end{figure}

 In this study, an \textit{in situ theranostic} platform was established allowing for interleaving diagnostics with magnetically focused MFH-therapy and image-based temperature monitoring without the need for target re-location. This \textit{theranostic} platform consists of a commercial preclinical MPI system (MPI 25/20FF, Bruker BioSpin MRI GmbH, Ettlingen) as a tomographic imaging device equipped with a custom-made hyperthermia insert \cite{wei2020} serving as a therapeutic tool. A powerful user-interface accesses the hardware communication and signal generation for the entire \textit{theranostic} platform. An MPI-based thermometry \cite{Stehning} feed-back-loop is included to comply with predefined temperature limits as well as to control hyperthermia duration - accounting for safety and efficacy. The user-interface allows to define all imaging and heating parameters, enabling selection of a therapy volume based on existing images. \
Before assessing the whole integrated MPI-MFH platform, the general capacity of MFH within the FOV was tested. SPION suspensions and water samples of the same volume were subjected to MFH and the temperature profile was monitored with a thermal camera (see figure \ref{fig:MPI guided MFH}C). Depending on the position within the sample holder, and therefore the relative position within the MPI-MFH platform, temperature increases of 15.6K up to 20.9K were observed for aqueous SPION samples. In the  control samples without SPION only negligible heating (up to 0.1K) was observed. We therefore conclude, that the source of the detected SPION suspension heating was caused  by the AMF induced MFH and did not result from system or phantom related heating effects.\
Subsequently, the extent of the field of therapy (FOT), i.e. the general working space allowing MFH and MPI , was determined by subjecting aqueous SPION samples distributed within the MPI-MFH platform to a non-specific global AMF (focus fields OFF). In the relative center of the MPI-MFH platform temperature increases of up to $\Delta {T}$= 22.6K were observed with a steady decline along the system's x-plane.  Eventually, AMF became ineffective at a distance of 8 mm from the system's center in both directions of the x-plane. We conclude that the implemented MPI-MFH platform offers a usable FOT of approximately 16x16x16 mm$^3$ (x,y,z) (see figure \ref{fig:globalMFH} ) while using global heating with this hyperthermia insert.

\begin{figure}[ht!]
\centering
\includegraphics[width=\linewidth]{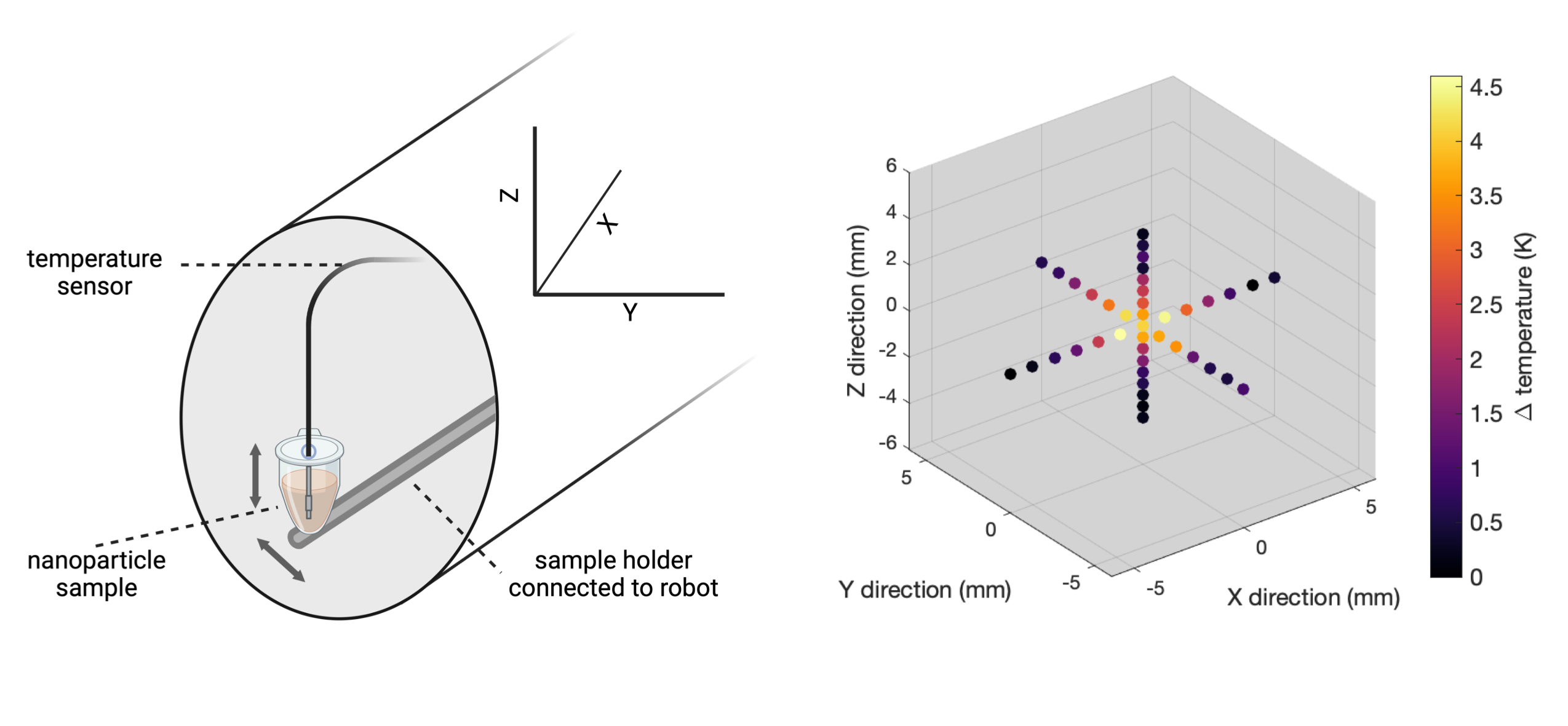}
\caption{\textbf{MFH-FFR} The system's FFR, presumably a point, was coordinate fixed to the relative center of the MPI-MFH platform. An \SI{8}{\micro\liter} SPION sample was subsequently moved along all three axes in 1 mm increments and subjected to AMF revealing the MFH's spread function relative to the FFR.}
\label{MFH-FFP}
\end{figure}

\subsection*{MPI-MFH theranostic platform offers a localized field of therapy}
In order to minimize off-target side-effects a focused application of hyperthermia is highly desired in most imaginable applications. Estimating the extend of the MFH target therefore provides crucial information on the minimal achievable therapeutic volume within our set up. With focus fields ON the highest temperature of +4.6 K was measured at the coordinate center  and predictably decreased with increasing distance along each axis with the FFR fixed (see figure \ref{MFH-FFP}). Corresponding to the greater drive field gradient strength in z-direction (42 mT in z- vs. 17 mT in each x- and y- direction), a steeper temperature distribution is observed in z- as compared to y-direction (see figure \ref{MFH-FFP}right). In x- and y- direction, the temperature increase becomes negligible beyond a distance of 5 mm to the FFR ($\Delta$ T = 0 K \& 0.7 K (+x \& -x), and $\Delta$ T = 0.5 \& 0.6 K (+y \& -y ). In z-direction, no significant temperature increase was observed beyond 3 mm to the FFR ($\Delta$ T = 0.5 K and 0.6 K for +z and -z respectively). We thus assess the heating function of our system's fixed FFR to affect a volume of 5x5x3 mm$^3$ (x, y, z) corresponding to the spatial distribution of the magnetic gradients within the hyperthermia insert.  

\subsection*{MFH  volume depends on the extent of focus and properties of the SPION solvent}
\begin{figure}[ht!]
\centering
\includegraphics[width=\linewidth]{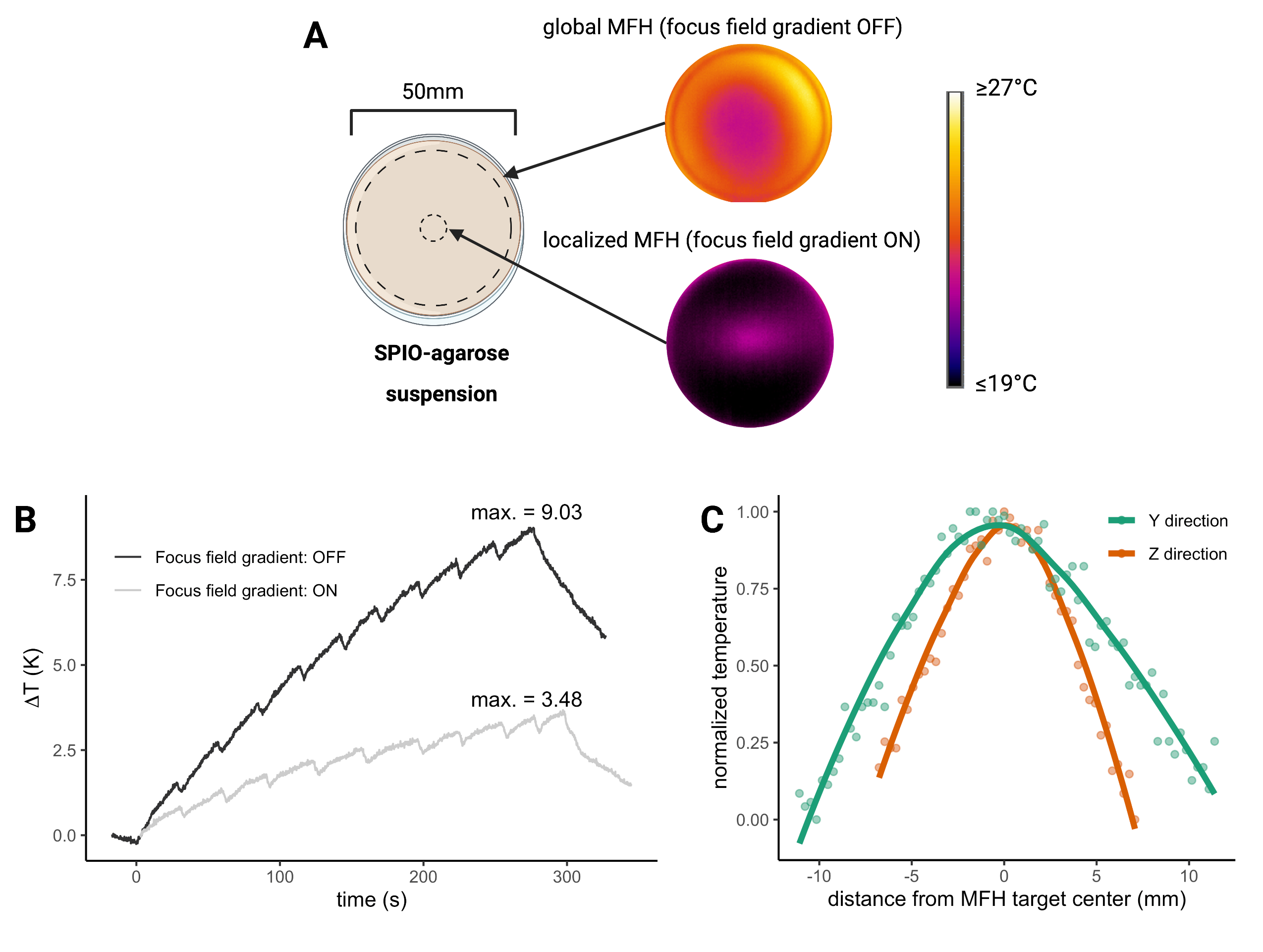}
\caption{\textbf{MFH in agarose} (A) Thermal camera monitored extent of temperature change in a SPION filled petri dish subjected to localized and global MFH. (B) Both corresponding temperature curves are shown versus time. (C) Due to the differences in focus field gradient strengths along the 3 orthogonal planes within the MPI-MFH platform, the spatial extent of heating differs for Y and Z axis.}
\label{fig:Agar}
\end{figure}

The actual temperature distribution inside a physical target is not only defined by the shape of above estimated affected volume, but by the heat conduction properties of said target as well. The temperature distribution will become due to intrinsic heat conduction more fuzzy than expected from above discrete points in air (see figure \ref{MFH-FFP}) .  To illustrate this effect, the surface temperatures of  a thin slice of gelled agarose ( 0.6\% w/w) containing 1mg(Fe)/ml SPION were recorded upon focused (focus field ON) and global (focus field OFF) MFH  (see figure \ref{fig:Agar}). 
During focused MFH  with the FFR at the center of the agarose-SPION sample, substantial temperature increase  in y-direction can be detected in the thermal picture beyond a distance of \SI{25}{mm} from the center of MFH application. In z-direction, the heated area extents to approximately 10mm in both directions from center. Assuming a similar extent of heating in x- and y-direction, the affected region exceeds \SI{50}{mm} x \SI{50}{mm} x \SI{20}{mm} in hydrogel (see figure \ref{fig:Agar}C). The heating function in a continuous, heat conducting sample is thus 10 times more blurred than expected from thermally independent sampling. \
Global heating of the entire sample, resulted in higher maximal temperatures  in the center of the sample as compared to local heating (see figure \ref{fig:Agar}B). During global heating, the entire SPION samples are heated while during local heating the heated spot (i.e. the FFR) is constantly dissipating heat to the cooler, unheated surroundings by conduction. 

\subsection*{Iron core size as predictor for heating efficiency in commercially available SPION}
\begin{figure}[p]
\centering
\includegraphics[width=\linewidth]{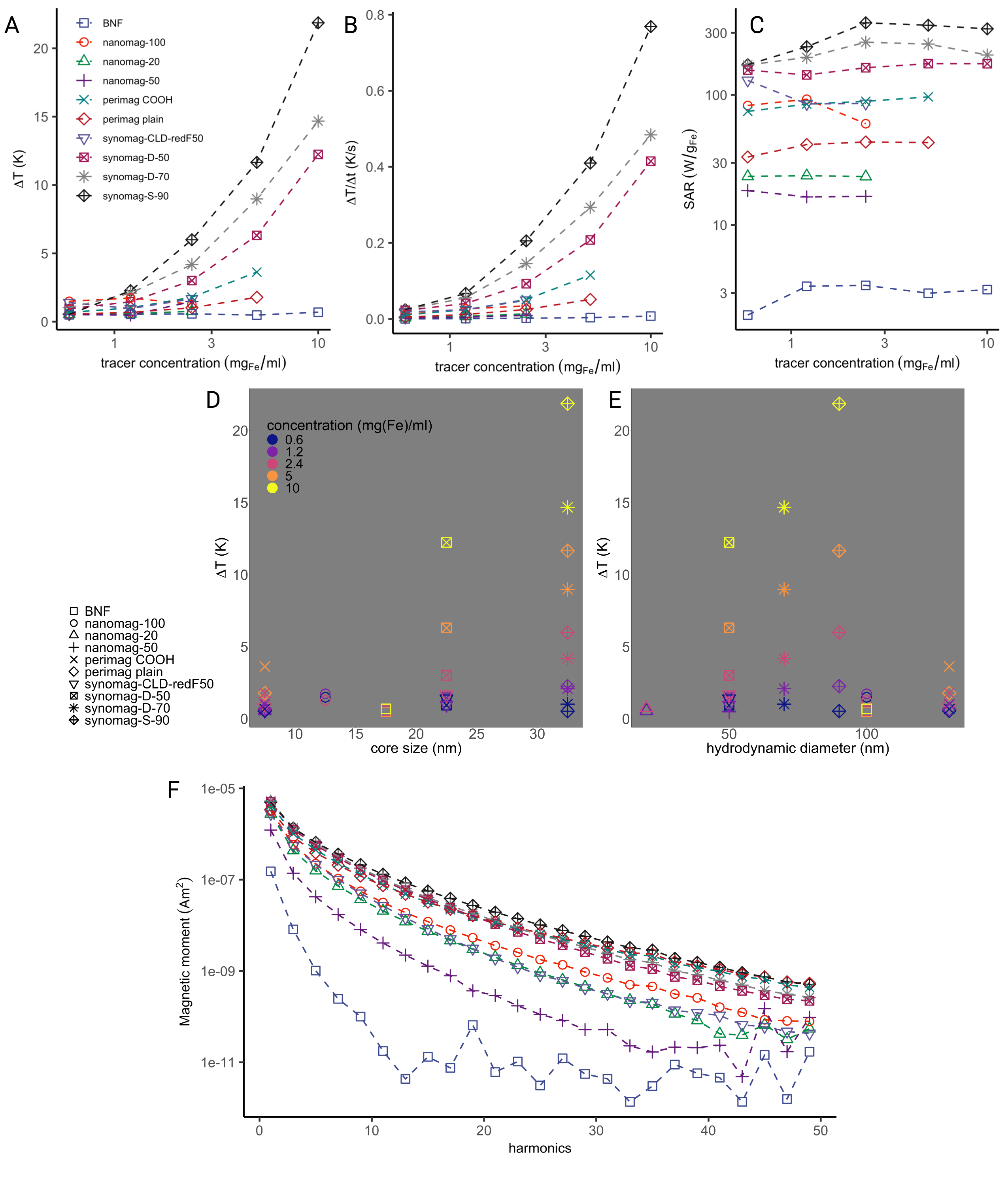}
\caption{\textbf{Comparing heating efficiency of SPION} Display of maximally achieved temperature increase (A), heating rates (B), SAR (C) for different SPION and concentrations during MFH application. Dependence of iron concentration and core size (D) or hydrodynamic diameter (E) on maximally achieved temperature increase during MFH application. Iron weighted magnetic moment of different SPION (MPS measured) (F).}
\label{fig:tracer}
\end{figure}

 In a theranostic application, both the image quality and therapeutic value (i.e. in our case heating efficiency) have to be considered. Therefore, we investigated the heating performance of 10 different SPION known for their good imaging quality. Only three of the SPION used in this study showed clear temperature increase upon MFH application. As expected, the maximal temperature showed a strong dependence on the iron concentration with an overall maximal temperature increase for each SPION being achieved at their respective commercial stock concentration (see figure \ref{fig:tracer}A). The best heating abilities were observed for  SPIONs at stock concentration (10mg(Fe)/ml): Temperatures of 37.2K, 27.2K, and 20.6K were measured for synomag-S-90, synomag-D-70 and synomag-D-50 respectively. The respective SPION's rate of heating corroborates the dependency on the iron concentration (see figure \ref{fig:tracer}B). 
The SPION's calculated SAR in aqueous suspension showed an initial increase with increasing iron concentration for synomag-D-70, synomag-S-90 and BNF, perimag plain, and perimag COOH. This was followed by plateaued or even slightly decreasing SAR values at higher iron concentrations. The remaining SPION displayed quite constant SAR values despite increasing concentration (see figure \ref{fig:tracer}C). 
When comparing the maximal temperature achieved during MFH in relation to the SPION iron core size, a dependence of heating efficiency on core size becomes evident. The highest temperature was achieved with the largest core size at the highest iron concentration (synomag-S-90 at 10mg(Fe)/ml,  see figure \ref{fig:tracer}D). 
Comparing the maximal temperature achieved during MFH in relation to the SPION hydrodynamic diameter a similar dependency is observed, however less pronounced. The highest temperature was achieved with the second largest core size at the highest iron concentration (synomag-S-90 at 10mg(Fe)/ml, see figure \ref{fig:tracer}E) while the SPION sample with largest hydrodynamic diameter (perimag plain, perimag COOH \diameter = 130nm) achieved only negligible temperature increase at highest concentration.\
The iron weighted magnetic moment of the different SPION showed a similar course compared to the maximally achieved temperature increase, slope and SARs of the samples (see figure \ref{fig:tracer}F). The lowest magnetic moments, and steepest decay with increasing harmonics was observed for BNF SPION. Highest temperature values were observed for synomag SPION.

\subsection*{MPI-MFH based theranostic platform offers spatial and temporal control of hyperthermia} 
\begin{figure}[!ht]
\centering
\includegraphics[width=\linewidth]{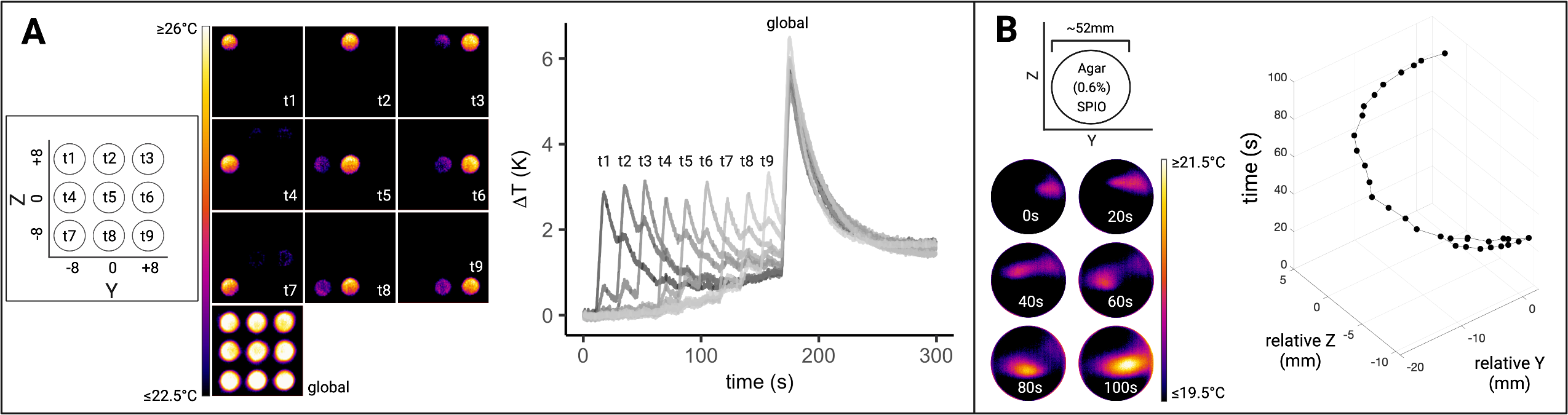}
\caption{\textbf{Illustration of MFH target control} (A) Thermal camera sequence of temporal and spatial control of MFH targeting via arbitrary FFR shifts (focus field ON): 9 discrete SPION samples were positioned horizontally at 9 locations within the MPI-MFH platform representing distinct MFH targets. SPION samples were heated successively starting with the sample located in the upper left corner (relative position = 0, -8, 8 mm for x,y,z), ending with global heating of all samples within the FOT as indicated by the individual temperature curves.  (B) MFH targeting in a continuous  agarose-SPION suspension. The MFH target describes a circle through the agarose-SPION hydrogel as displayed by the thermal camera snapshots at different representative time-points on the left. right: Spatio-temporal distribution of the maximal temperature on the agarose surface every 10s corroborates a circular MFH trajectory.}
\label{fig:directional}
\end{figure}

In order to avoid systemic and/or off-target side-effects, spatial control and local confinement of heat application during hyperthermia treatment is highly desired. By adjusting the focus field gradients in x-, y- and z- direction samples at different, disjunct positions can be targeted within one sequence. To investigate that capability, several SPION samples were placed at different positions in the yz- plane (x = 0) and by adjusting the corresponding gradients, arbitrary spatial selection of MFH  was demonstrated (see figure \ref{fig:directional}). Thermal camera images show snapshots of temperature maps at different time points while independent targets were aimed at. The adjacent plot indicates the extent of heating relative to the start temperature. The results display a sequential MFH-walk through each sample position, ending with a global application of AMF.\
The same approach has been applied to a homogeneous agarose-SPION mixture placed in a petri dish such that the MFH target described a circular trajectory in the SPION-laced hydrogel. Next to the corresponding thermal camera images at selected time points, the local temperature maxima within 10 s intervals are plotted. The trajectory covering the extent of the agarose area is plotted further illustrating the effectivity of the localized sequence for heating.

\subsection*{MPI-MFH theranostic platform allows for tomographic imaging, localized hyperthermia and multicolor thermometry} 
\begin{figure}[!ht]
\centering
\includegraphics[width=\linewidth]{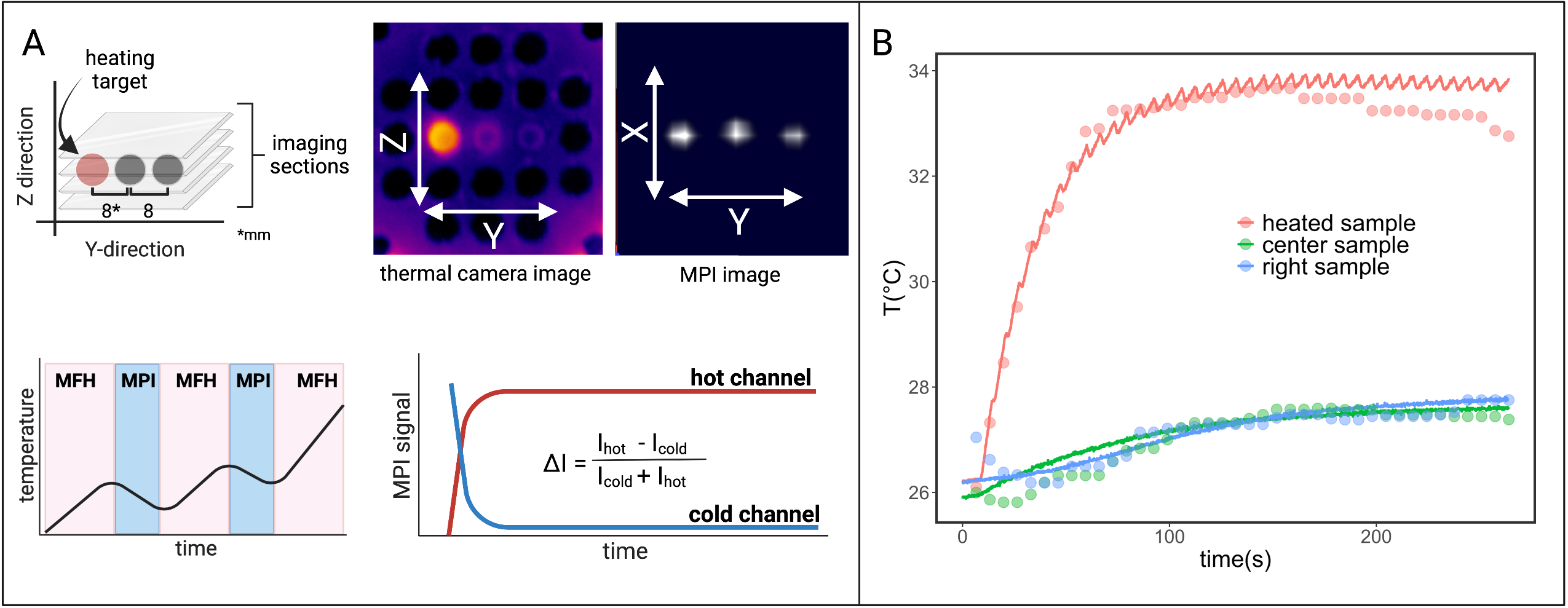}
\caption{\textbf{Interleaved MPI-MFH} Three SPION samples at equidistant positions within the MPI-MFH platform (x,z = 0 mm; y = -8, 0, 8 mm form center) were subjected to alternating MPI-MFH sequences. By adjusting the MFH-FFR, the SPION sample positioned at y= -8 mm, was targeted by MFH while MPI images were acquired of all samples (A). A sketch of the MPI-MFH sequence is depicted in the lower left. The MPI images are reconstructed applying the multi-color method for MPI-based thermometry (lower right).
 Multi-color reconstructed temperature values (red, green and blue dots) are overlayed with thermal camera measurements (sampling rate of approximately 10Hz (red, green and blue lines)) (B).}

\label{fig:MPI-MFH1}
\end{figure}

To testify the successful integration of a true theranostic device, consisting of an imaging and an interventional module, a sequence of alternating MPI and MFH episodes was applied to several SPION samples similar to the experimental set up used above. This time, an additional feature was added to the platform by utilizing the multi-color method for MPI based thermometry. Here the temperature distribution is retraced during MFH using specific image reconstruction of the MPI images acquired during the MPI-MFH experiment.
The multi-color reconstructed temperature values were subsequently validated by the temperature profile of  thermal camera recordings. The resulting temperature profile, overlayed with the MPI-based temperature values, are presented in figure \ref{fig:MPI-MFH1}B. \\
The left hand SPION-sample (y = -8 mm) was subjected to MFH and consequently shows a steep temperature increase upon application onset while the temperature of the 2 adjacent samples remained relatively constant as compared to baseline.\
Since the absolute MPI-reconstructed temperature values are based on normalization with the temperature values of the thermal camera measurements, absolute maximal temperature values are shared between these two modes of temperature measurement. However, comparison of the maximal relative temperature increase ($\Delta{T}$) as well as qualitative comparison of the progression of heating between thermal camera and MPI based temperature may be effective for an interpretation of the results. 
The MFH-heated sample showed a maximal temperature increase measured by thermal camera of 7.8 K while a maximal increase of 7.9 K was obtained with the multicolor MPI-method. The center sample's temperature (y = 0 mm) increased 1.8 and 1.7 K for thermal camera and MPI based temperature values respectively. The largest difference, could be observed for the sample with the furthest distance to the heating target (y= 8 mm) where 1.7 K (thermal camera) and 0.8 K (MPI reconstructed) increase in temperature was measured. The rapid initial heating of the MFH targeted sample can be similarly well depicted with thermal camera as with the MPI multicolor method. Subsequently, there is a slight overshoot of the MPI reconstructed temperature values as compared to thermal camera measurements followed by a brief period when both temperature values plateau at similar values. 

\subsection*{MPI-MFH platform adjusts to new hyperthermia target during application}

\begin{figure}[!ht]
\centering
\includegraphics[width=\linewidth]{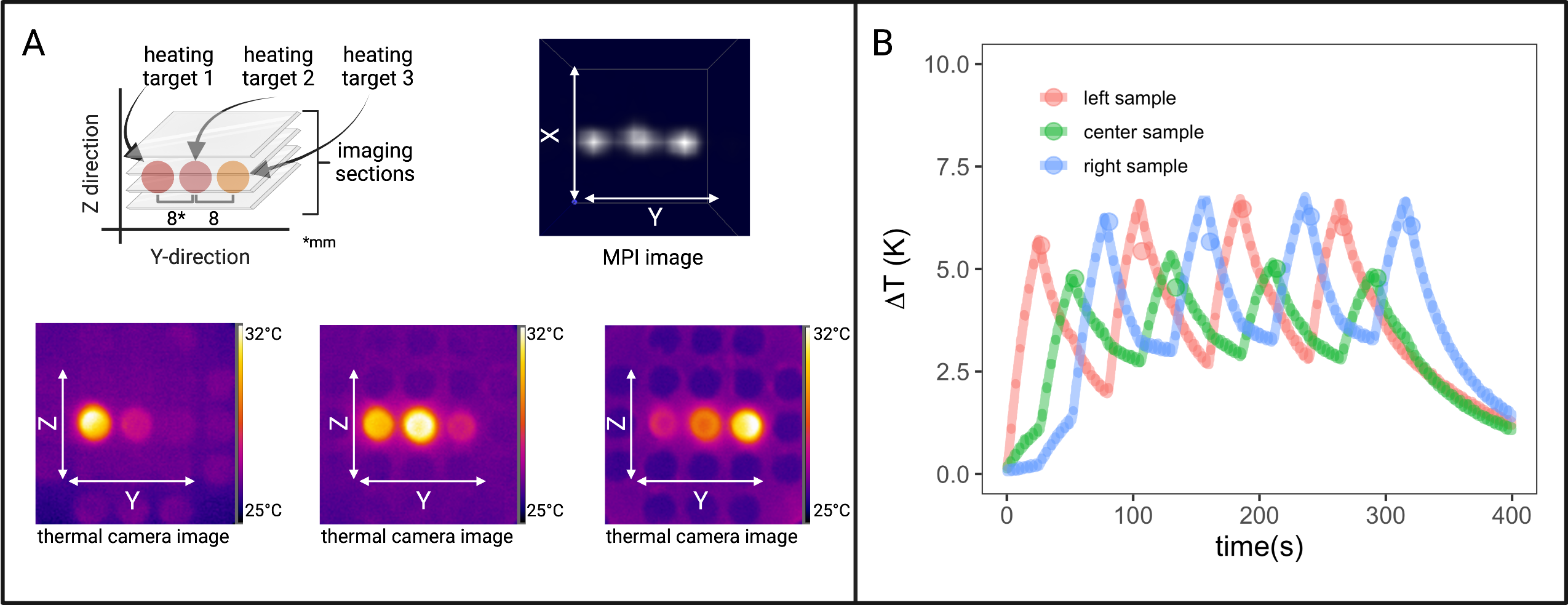}
\caption{\textbf{Interleaved MFH and MPI with alternating MFH targets}  Three small SPION samples at equidistant positions within the MPI-MFH platform (x,z = 0 mm; y = -8, 0, 8 mm form center) were subjected to alternating MPI-MFH sequences. By adjusting the MFH-FFR, the SPION samples were targeted successively while the temperature profile was monitored with a thermal camera (A).
 Multi-color reconstructed temperature values (red, green and blue dots) overlayed with thermal camera measurements (sampling rate of approximately 10Hz (red, green and blue lines)) (B).}

\label{fig:MPI-MFH with alternating heating targets}
\end{figure}

The previous two sections showed the feasibility of MFH-target localization and alternating MPI-MFH sequences  with MPI based thermometry in one theranostic platform. In the following, we present the results of the combination of both, i.e. a MPI-MFH platform with arbitrarily alterable theranostic targets (see figure \ref{fig:MPI-MFH with alternating heating targets}). The thermal camera measurements as well as the plot showing the heating process over the duration of the experiment (see figure \ref{fig:MPI-MFH with alternating heating targets}B, red, blue \& green lines), show the successive heating of three small SPION samples beginning with the left hand sample. The MPI images acquired during MPI-MFH sequences (see figure \ref{fig:MPI-MFH with alternating heating targets}A upper right) served as basis for subsequent MPI-based multi-color thermometry. The resulting calculated temperature values show adequate overall agreement when aligned with the thermal camera recordings (see figure\ref{fig:MPI-MFH with alternating heating targets}B; red, green \& blue dots).

\subsection*{MPI-MFH procedure is limited in circulatory system}
\begin{figure} [!ht]
\centering
\includegraphics[width=\linewidth]{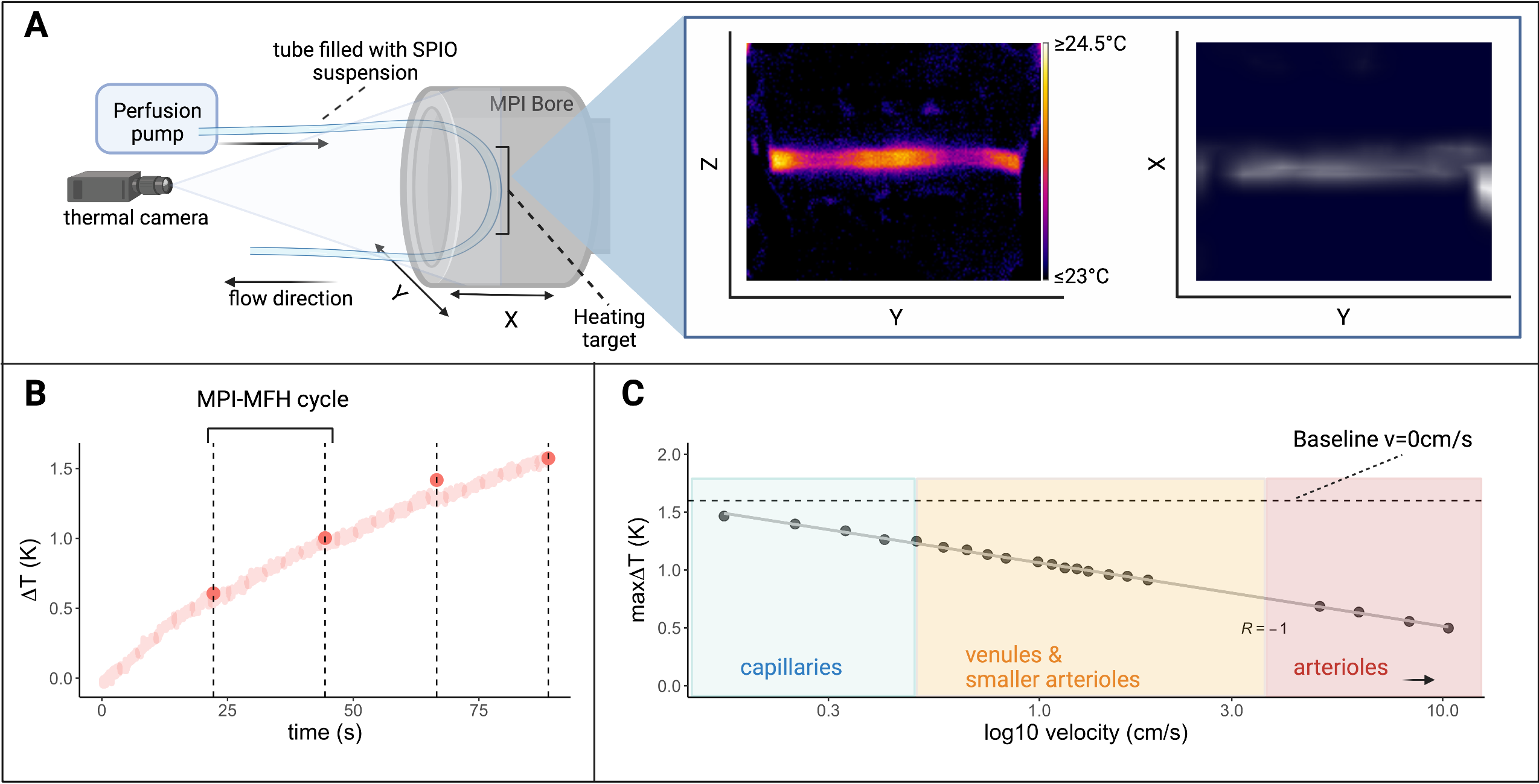}
\caption{\textbf{Effect of MFH in a flow phantom} A flow system connected to a tube filled with a SPION was placed inside the MPI bore to mimic (blood) circulation and investigate the effect of circulation velocity on heating efficiency. A thermal camera was placed in front of the MPI bore such that the heating target (inside the AMF of the hyperthermia insert) as well as portions of the circulation tube outside the AMF were visible. The temperature progression outside the AMF was defined as offset temperature and subtracted from the heating target inside the MPI bore for each timepoint during heating. The mean temperature increase of a 24s heating period was plotted for different flow velocities. The temperature increase at a circulation velocity of \SI{0}{\cm\per\second}. Flow velocities of different blood vessel types \cite{textbook} were approximated with the results and a linear regression model was fitted to the data points.}
\label{fig:dynamic}
\end{figure}

This study aims to testify the general characteristics of an MPI and MFH-based theranostic platform. As most reports on treatment with MFH utilized accumulated or injected SPION \cite{Tay,Kuboyabu}, stationary phantoms with well defined SPION samples seemed to be adequate test objects. However, systemic SPION applications by i.v. injection are not well represented by static phantoms and the application gap to living breathing beings has still to be tackled. In order to  narrow down that gap a little, we included one very important parameter into our phantoms: Forced transport of SPION solutions through the MFH-target volume similar to blood flow reduced interaction and heating times.   
A circulatory tubing system in plain sight of the thermal camera was set up within the bore of our MPI-MFH platform. Without flow (v0= \SI{0}{ml/min}) a maximal temperature increase of \SI{1.6}{K} was achieved on the surface of the thin tubing. Flow velocities of v=\SI{0}{\cm\per\second} up to v=\SI{10.3}{\cm\per\second} were applied achieving temperature increases of 1.6 and 0.5K respectively. An exemplary MPI image of the circulation tube (v = \SI{0.65}{\cm\per\second}) and the corresponding thermal camera image during heating is depicted in figure \ref{fig:dynamic} upper right. MPI-based multi color reconstruction of the temperature profile during MPI-MFH application was attempted for each of the indicated flow velocities. However, only reconstruction of the slowest velocities (0, 0.17, 0.25, 0.33, 0.41 cm/s ) showed good overall alignment with the temperature readings from the thermal camera (shown in the exemplary temperature progression and multi-color reconstructed temperature values in figure \ref{fig:dynamic} lower left). At higher velocities, such alignment was found to be inconsistent.  It has to be stated, that the small temperature increase seen on the tubing walls by thermal camera were achieved by very short AMF in  \SI{10}{\milli\liter} tube volume as compared to the \SI{140}{\micro\liter} volume of the glass vials with SPION optimized for imaging, but not for highest possible heating properties. 

\section*{Discussion}

Theranostic platforms have gathered quite some interest, on the one hand for minimizing invasiveness and off target effects during and post-therapy and on the other hand for augmenting the real-time reactivity during intervention. Exploiting MPI imaging properties as a mean of generating temperature increase for potential therapeutic interventions has been reported in previous works \cite{Wells2020,Murase2015}. Indeed, MPI-MFH based platforms have been described for targeted MFH delivery\cite{Malhotra2019,Tay,behrends2019selfcompensating}.\ 
However, interleaving MPI and MFH sequences, offers the potential for a therapeutic feed-back loop by periodically imaging the targeted SPION samples within the FOT which, by means of MPI based thermometry, provides potentially real-time monitoring and therapy-parameter adjustments. 
Purpose of this study was to demonstrate the general feasibility of an MPI based theranostic platform which entails tomographic imaging, thermometry and directional, highly localized "therapy" within the same system in the context of phantom experiments. We were able to demonstrate high spatial control on the scale of millimeters for both imaging and MFH application and adequate resolution of temperature reconstruction \textit{in situ}. 
MPI based multi-color reconstruction of temperature values depends on the linear relationship of imaging signal and SPION temperature. The highest discrepancy between multi-color reconstructed  and thermal camera monitored temperature values were observed for high velocities in a circulation phantom. 
We conclude therefore, that the limit of multi-color temperature reconstruction is reached here and based on the comparatively slight temperature increase achieved during MFH of high velocity circulation. Longer imaging sessions with more averages or improved SPION could be used to enhance the multi-color reconstruction result. It should be noted here, that the thermal camera measures the temperature of the tubing's surface, while MPI based temperature reconstruction is based on the  SPION's temperature.
We observed good MFH properties for SPION also exhibiting high measured magnetic moments. This may indicate that SPION properties improving imaging quality, such as their shape and size, impact MFH performance and has to be considered while pursuing theranostic SPION. 
\section*{Methods}

\subsection*{MPI-MFH theranostic platform}

The commercial MPI system used in this study generates a magnetic field gradient of up to 2.5 T/m in z-direction and 1.25 T/m in x- and y- direction (Selection Field) used in our case for both spatial encoding during imaging and definition of the heating region in the MFH-based intervention. By means of 3 orthogonal homogeneous magnetic offset fields (Focus Fields) with amplitudes of up to 17 mT, 17 mT and 42 mT for x-, y-, and z-direction respectively (see figure \ref{fig:MPI guided MFH} A for a representative overview), both the imaging FOV and the field of therapy (FOT) can be shifted in space. During MPI image generation, the SPION are excited using 3 time-variant orthogonal homogeneous magnetic fields (Drive Fields) with amplitudes of up to 14 mT in each direction and resonance frequencies of 25 kHz. For heat generation during MFH application, the SPION are excited using a custom-made single-axis time-variant homogeneous magnetic field (Hyperthermia Field) applied by a hyperthermia insert \cite{wei2020} which was designed as a gradiometric coil to decouple it from the transmit-receive Drive Field coil set (for imaging). The hyperthermia insert is placed concentrically inside the MPI's bore. A radio frequency power amplifier (T\& C Power Conversion, Inc. Model AG 1012 Amplifier Generator) was used to generate and amplify the signal of the system. When applying the maximum working power of 600 W, the hyperthermia insert is capable of generating an average field strength of \SI{11.2}{mT}, with a standard deviation of $4\times 10^{-4}$ mT over a volume of 22.5 mm $\times$ 18 mm $\times$ 12 mm (for x,y,z) around the center of the heating coil (see figure \ref{fig:MPI guided MFH}B for a representative overview). The MFH field strength for all measurements presented in this work was set to 10 mT along the x-axis. In order to limit the power transferred to the MPI scanner, the hyperthermia insert has compensation winding which create a magnetic field in the opposite direction of the primary heating field \cite{wei2020}. The hyperthermia insert itself is cooled with oil to avoid hardware overheating and additionally an air flow system was applied at the MPI bore entry. This was of systemic importance for achieving maximal heating durations and steady temperatures of the targeted SPION, while minimizing the risk of overheating. As the MFH-excitation frequency of 715 kHz lays within the imaging detection bandwidth of the MPI its low noise receive amplifiers were safe guarded with an additional low-pass filter and/or amplifier blanking device. While the low-pass filter suppresses not only the MFH feed-through-signal but also the high frequency components for imaging purposes, the latter implementation has no negative impact to the imaging bandwidth but needs active components which interacts with the theranostic scan program. With the hyperthermia insert installed, the free accessible bore diameter of the integrated theranostic platform measures 65 mm, suitable for preclinical \textit{in vivo} studies. Since MPI imaging and MFH therapy are both based on magnetic SPION excitation, several measures needed to be implemented to prevent harmful subsystem interactions.

\subsection*{Temperature measurements}
During experiments, if not specifically stated otherwise, temperature profiles were recorded using a thermal camera (FLIR Systems, A8303sc with custom macro lens, 20 mK temperature resolution and 30 Hz frame rate and 308 $\mu$m pixel size) that was placed approximately \SI{1.5}{m}in front of the MPI scanner system with unobstructed line of view inside the bore. Temperature progressions were analyzed by manually defining ROIs covering the respective sample and extracting the maximal temperature per time point. 

\subsection*{MFH applications in SPION phantom}
If not stated otherwise, all temperature measurements during MFH application were done in a 3D printed circular sample holder (\diameter\SI{59}{\milli\meter}, L = \SI{48}{\milli\meter}) designed to fit the hyperthermia insert and equipped with several notches to allow for sufficient air flow (see figure \ref{fig:MPI guided MFH}). The sample holder can accommodate 9 samples at once, spaced 2 mm apart in y- and z- direction. Thin walled glass tubes (6 $\times$ 50 mm , Disposable culture tubes, Borosilicate Glass, Kimble) served as punctiform MFH targets. The glass tubes were sealed off with paraffin to avoid drying out. The glass tubes were placed horizontally inside the sample holder, the bottom facing the thermal camera such that a circular target is presented, resulting in a 3x3 grid spanned in y-z- direction between -8 and 8 mm relative to the hyperthermia insert's center for any given location on the x plane. \\
As proof of concept, the sample tubes filled with 140 µl of SPION suspension (synomag-S-90, 10mg(Fe)/ml) or 140 µl of water, were heated in 2 separate MFH sessions using the maximal power of 600W with 2 heating cycles of \SI{22}{\second}.

\subsection*{Characterization of field of therapy}
Upon equipping a sample holder (see figure \ref{fig:MPI guided MFH}C) with SPION samples positioned at all locations in y- and z- direction, the entire sample holder was moved through the hyperthermia insert in x- direction in seven \SI{8}{mm} increments going from -24 mm to +24 mm relative to the hyperthermia insert center. Locations in x-direction are hereby corresponding to the center of the SPIONs sample. At each incremental position, MFH was applied to the entire inner volume of the hyperthermia insert (global MFH) yielding a total of 63 individual positions and covering an area of 48mm $\times$ 16mm $\times$ 16mm (x, y, z). A sketch of the experimental set up is depicted in figure\ref{fig:globalMFH}. MFH was applied using the maximal power of the hyperthermia insert (600 W) with 2 heating cycles of 25s each.

\subsection*{3D Characterization of localized FFP-MFH}
To precisely determine the system's heat production function, i.e. the extent of SPION heating along the orthogonal planes of the MPI-MFH experimental set up, the location of the FFR was fixed at the center of the hyperthermia insert. A punctiform calibration sample was attached to a holder rod and moved to each position inside the MPI-MFH platform's FOT in 1 mm increments (see figure \ref{MFH-FFP}left), very much alike the process of system matrix acquisition. The calibration sample consisted of a temperature-stable PVC tube filled with \SI{27}{\micro\liter} SPION suspension (synomag-D-70, plain, Micromod Germany, 10 mg(Fe)/ml). Heating was achieved by sending power through the heating coil (P = 300 W, one cycle of 30 s) at each spatial location. Focused MFH was achieved by focus field gradients in x,y,z direction of 1.25 T/m , 1.25 T/m , 2.5 T/m respectively. Temperature of the SPION sample was monitored by a commercial fiber optic temperature sensor (TS2, Weidmann-Optocon, Germany) introduced through the lid of the PVC tube, with ample cooling time between different spatial positions. 

\subsection*{Characterization of localized and global MFH in a continuous hydrogel sample}
A SPION (synomag-D-70) suspension was added to molten agarose (0.6\% by weight; simulating brain tissue visco-elasticity\cite{Chen:2004vj}) leading to a concentration of 1 mg(Fe)/ml in a culture dish (glass bottom dish 50mm x 7mm, Wilco Wells). Upon solidifying to an approximately 5 mm thick layer, the dish was placed upright in the center of the hyperthermia insert with the lid off the dish, such that the circular surface of uncovered agar faced the thermal camera for high speed thermal monitoring. Thereby, heating of the thin agar-layer could be observed in y and z-direction on the surface of the hydrogel. The SPION-agarose suspension target was heated (P = 600W) either locally (focus field gradients of 1 T/m , 1 T/m , 2 T/m (x,y,z) with the FFR located at the center of the MPI-MFH system or globally (focus field off i.e. no FFR). An overview of the experimental set up is shown in figure \ref{fig:Agar}A.
 For comparison of the maximally achieved temperature during both measurements, a circular ROI was manually defined at the center of the thermal camera monitored culture dish, and the maximal temperature was extracted over time. The extent of heating during localized MFH was determined by placing 2 orthogonal lines of the same size through the center of the monitored agarose suspension and extracting the temperature values along those lines which correspond to y- and z- direction inside the MPI-MFH platform. figure \ref{fig:Agar}B).

\subsection*{Targeting the field of therapy}
The selection field gradient defines the spatial resolution during MPI\cite{Rahmer:2009uh} by determining the FFR. Simultaneously, the magnetization response of the deployed SPION influences to the spatial resolution: A steep magnetization slope leads to higher spatial selectivity.\\ 
A similar mechanism is expected to apply to the MFH targets. Therefore, component parameters of the MPI system define the extent of the FOT during AMF applications and the point spread function of the FFR defines the focus of the MFH treatment since only SPION within the FFR are susceptible to time varying changes. Finally, any shift to the FFR using focus fields allows to similarly shift the FOT to any spatial target within the MFH FOT.
\\
The above described sample holder ( see figure \ref{fig:MPI guided MFH}C) was used to demonstrate directional control of heat application. The phantom was equipped with 9 SPION samples (synomag-S-90, 5mg(Fe)/ml, Micromod Germany) and placed at the center of the hyperthermia insert. Subsequently, each of the sample locations (in y-and z-direction) was targeted successively using the same MFH parameters for each location: P = \SI{600}{\watt}, focus field gradient = \SI{1}{T/m}, \SI{1}{T/m}, \SI{2}{T/m} (x,y,z), 1 heating cycle of duration = \SI{10}{\second}.

\subsection*{Comparison of different SPIONs} 	
\begin{figure}[ht!]
\centering
\includegraphics[width=\linewidth]{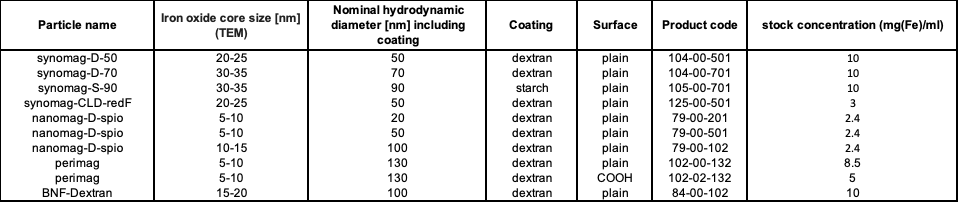}
\caption{Table of used SPIONs}
\label{fig:tracer table}
\end{figure}

To keep things simple SPION entailing both good imaging and MFH quality would be best suited for optimal theranostic performance. We compared different commercially available SPION, which have been predominantly used for imaging purposes, concerning their ability to generate heat using the set up introduced here. A table of the SPION's characteristics can be found in (figure \ref{fig:tracer table}). 
The described SPION sample holder (see fig \ref{fig:MPI guided MFH}C ) was used. The heating performance of the different SPION samples was determined in successive separate MFH sessions at the same specifically chosen spatial locations to account for any potential phantom or gradient related inhomogeneity. Different SPION sample concentrations were used (0.6, 1.2, 2.4 mg(Fe)/ml for all, 5 and 10 mg(Fe)/ml when applicable).  The individual specific commercial stock concentration served as upper concentration limit for the respective SPION (see table \ref{fig:tracer table}). 
During MFH the maximal power of the hyperthermia insert (P= \SI{600}{\watt}) was applied for 2 consecutive heating cycles with 20s duration each. The thermal camera monitored temperature profiles were analyzed by extracting the maximal temperature per time point. The specific absorption rate (SAR) of each sample was then calculated from the temperature profiles using the following formula:
\begin{equation}
SAR = \frac{C_{H_20} \times m_{H_20}} {{m_{Fe}}} \times \frac{\Delta{T}}{\Delta{t}}
\end{equation}
Here $C_{H_20}$ is the heat capacity of water (\SI{4.168}{J/gK}), $m_{H_20}$ is the mass of water of the sample, ${m_{Fe}}$ is the mass of iron in the sample, and $\frac{\Delta{T}}{\Delta{t}}$ is the slope of temperature increase during the first time points of the MFH application. \\

MPS measurements (pure devices, drive field = 25mT, excitation frequency = 25kHz) were performed using \SI{50}{\micro\liter} of each SPION sample (see \ref{fig:tracer table}) at their respective stock concentration. For comparison, the results were normalized for the different iron concentrations. 

\subsection* {Interleaved MPI-MFH and MPI thermometry}
The above described sample holder (see fig \ref{fig:MPI guided MFH}C ) equipped with three SPION samples was placed inside the MPI-MFH system such that the center of each of the samples located at x, z=0 was aligned in y-direction at positions y = +8 mm, 0 mm \& -8 mm (see figure \ref{fig:MPI-MFH1}A).
The three samples were periodically imaged (MPI). The imaging sequence was hereby alternating with the MFH-sequence, targeting the SPION sample at y = -8 mm (see \ref{fig:MPI-MFH1}). MPI-MFH sessions consisted of blocks lasting 6.1s per repetition (1.1 s (50 averages) imaging and 5s heating respectively), with a total number of 40 repetitions (acquisition time 248s). Other parameters were: acquisition bandwidth of 1.25MHz, drive field strength of 14, 14, 14 mT (x,y,z), selection field gradient strength of 1, 1, 2T/m (x,y,z), and drive field FOV of 20, 28, 14 mm (x,y,z). Images were acquired with  matrix size = 20, 20, 20 (x,y,z). The FOV was set, analog to the FOT, to 40, 40, 20mm (x, y, z).\\ 
The acquired images were subsequently used for MPI based thermometry utilizing the "multi-color" method \cite{Stehning,Buchholz2022}: 
MPI thermometry is based on the calibration  of the MPI system to differently tempered SPION samples. According system matrices are acquired defining the temperature range of the SPION that can be detected within an imaging session. Here the lower base temperature of the sample ("cold channel") and the maximal achieved temperature ("hot channel") were chosen as lower and upper boundaries. As the MPI signal and the SPION temperature is linear dependant, temperature profiles of  samples in a given imaging session can be obtained by reconstructing the MPI signals using both hot and cold channels \cite{Stehning}. 
Following the multi-color method, two system matrices were acquired at temperatures that were assumed to represent the upper (\SI{43}{\degreeCelsius}) and lower (room temperature (\SI{23}{\degreeCelsius})) boundaries of the imaging session. For that, a water bath tempered SPION sample (synomag-S-90, Micromod Germany, 5mg(Fe)/ml) was placed in a 3D printed sample holder mounted on the computer controlled sample rod used for system matrix acquisition in our set up . 
MPI images were reconstructed with the two system matrices representing base and maximal temperature using the following reconstruction parameters: relative regularization value: 0.01; number of iterations: 15.\\

 
 In an experimental set up analog to the one described above, the objective was expanded to include alternating MFH targets during sequences of MPI-MFH application. Starting with the SPION sample located at the relative position y=-8mm, the samples were targeted in succession (y=-8mm, y=0mm,y=8mm) while images of all three samples were acquired for the duration of the measurement (see \ref{fig:MPI-MFH with alternating heating targets}). Reconstruction of MPI-based temperature values was accomplished using the same system matrices and reconstruction parameters as described in the previous section. The spatial resolution of the temperature maps originate directly from the imaging parameters.

\subsection*{Variations of flow velocity during MPI-MFH}
To study the effect of flow velocity on heating efficiency within the MPI-MFH platform, 10 ml of a SPION (synomag-D-70)- suspension at a concentration of 5mg(Fe)/ml were loaded to a flow system driven by a perfusion pump (PPS2, Multichannel Systems, Harvard Bioscience Inc.) connected to a loop (loop length= 375cm, $\oslash_{inner}$ = 0.2mm, $\oslash_{outer}$ = 0.3mm, silicone tubing, Tygon\textsuperscript{\textregistered} ) (see figure \ref{fig:dynamic}). The tube was placed at the relative position of x=0 such that the tubing segment serving as the theranostic target covered almost the entire length of the MPI bore in y-direction while the two ends of the tube were guided along a 3D printed holder to the pumping device outside the MPI system. Therefore, the relative spatial location of MPI-MFH target was at x, z = 0, spanning the entire diameter of the MPI's bore in y- direction. The experimental set up is depicted in figure \ref{fig:dynamic} upper left. \\
An alternating sequence of MFH and MPI (4 block repetitions of 21s MFH and 1s imaging) was used to heat and image at selected flow velocities the horizontal segment of the tubing. The maximal temperature increase achieved during global heating (600W) was analyzed for different flow velocities and put in relation to the approximate velocities of differently sized blood vessels \cite{textbook}.\\
The set up allowed for circulation of the SPION suspension in 0.1ml/min increments. The SPION suspension was pumped at different flow rates (0, 0.2, 0.4, 0.5, 0.6, 0.7, 0.8, 0.9, 1, 1.2, 1.3, 1.4, 1.5, 1.6,1.7, 1.8, 1.9, 2.25, 6, 10, 12 ml/min) corresponding to circulation velocities of 0.17, 0.25, 0.33, 0.41, 0.50, 0.58, 0.66, 0.74, 0.83, 1.00, 1.07, 1.16, 1.24, 1.32, 1.49, 1.65, 1.86, 4.96, 6.20, 8.26, 10.33 cm/s respectively. 


MPI images were acquired with the following parameters: matrix size = 15, 15, 5 (x,y,z),  FOV (analog to FOT) = 30, 30, 15 mm (x,y,z), drive field strength = 14, 14, 14, mT (x,y,z), focus field gradient strength = 1, 1, 2 T/m (x,y,z), drive field FOV = 28, 28, 14 mm (x,y,z) , bandwidth = 1.25MHz. 

MPI based multi-color reconstruction following the method described in section \textit{Interleaved MPI-MFH and MPI thermometry} was achieved with the "cold channel" acquired at 23°C and the "hot channel" acquired at 27°C using the same respective parameter values as during image acquisition. The number of measurement averages was set to 50 resulting in a total acquisition time of approximately 7h 44m. 
MPI images were subsequently reconstructed with the two system matrices using a  relative regularization value of 0.01 and 15 iterations. 
Thermal camera pictures have to be considered lower bounds for SPION temperature, as the tubing's silicoen wall may act as insulator obscuring the inner temperature from the outside wall. 

\bibliography{sample}

\begin{thebibliography}{10}
\urlstyle{rm}
\expandafter\ifx\csname url\endcsname\relax
  \def\url#1{\texttt{#1}}\fi
\expandafter\ifx\csname urlprefix\endcsname\relax\def\urlprefix{URL }\fi
\expandafter\ifx\csname doiprefix\endcsname\relax\def\doiprefix{DOI: }\fi
\providecommand{\bibinfo}[2]{#2}
\providecommand{\eprint}[2][]{\url{#2}}

\bibitem{Simon2021}
\bibinfo{author}{Simon, J.}
\newblock \bibinfo{journal}{\bibinfo{title}{Disease diagnosis and treatment;
  could theranostics change everything?}}
\newblock {\emph{\JournalTitle{Med Health Care Philos}}}
  \textbf{\bibinfo{volume}{24}}, \bibinfo{pages}{401--408},
  \doiprefix\url{10.1007/s11019-021-10015-6} (\bibinfo{year}{2021}).

\bibitem{Wiesing2019}
\bibinfo{author}{Wiesing, U.}
\newblock \bibinfo{journal}{\bibinfo{title}{Theranostics: Is it really a
  revolution? evaluating a new term in medicine.}}
\newblock {\emph{\JournalTitle{Med Health Care Philos}}}
  \textbf{\bibinfo{volume}{22}}, \bibinfo{pages}{593--97},
  \doiprefix\url{10.1007/s11019-019-09898-3} (\bibinfo{year}{2021}).

\bibitem{Gilham2002}
\bibinfo{author}{Gilham, I.}
\newblock \bibinfo{journal}{\bibinfo{title}{Theranostics - an emerging tool in
  drug discovery and commercialisation.}}
\newblock {\emph{\JournalTitle{Drug Discovery World}}}
  \textbf{\bibinfo{volume}{Fall}}, \bibinfo{pages}{17--23}
  (\bibinfo{year}{2002}).

\bibitem{Gomes2020}
\bibinfo{author}{Gomes~Marin, J.~F.} \emph{et~al.}
\newblock \bibinfo{journal}{\bibinfo{title}{Theranostics in nuclear medicine:
  Emerging and re-emerging integrated imaging and therapies in the era of
  precision oncology.}}
\newblock {\emph{\JournalTitle{Radiographics}}} \textbf{\bibinfo{volume}{40}},
  \bibinfo{pages}{1715--40}, \doiprefix\url{10.1148/rg.2020200021}
  (\bibinfo{year}{2020}).

\bibitem{Janib2010}
\bibinfo{author}{Janib, S.~M.}, \bibinfo{author}{Moses, A.~S.} \&
  \bibinfo{author}{MacKay, J.~A.}
\newblock \bibinfo{journal}{\bibinfo{title}{Imaging and drug delivery using
  theranostic nanoparticles.}}
\newblock {\emph{\JournalTitle{Adv Drug Deliv Rev}}}
  \textbf{\bibinfo{volume}{62}}, \bibinfo{pages}{1052--63},
  \doiprefix\url{10.1016/j.addr.2010.08.004} (\bibinfo{year}{2010}).

\bibitem{Herynek2021}
\bibinfo{author}{Herynek, V.} \emph{et~al.}
\newblock \bibinfo{journal}{\bibinfo{title}{Maghemite nanoparticles coated by
  methacrylamide-based polymer for magnetic particle imaging}}.
\newblock {\emph{\JournalTitle{J Nanoparticle Res.}}}
  \textbf{\bibinfo{volume}{23}}, \bibinfo{pages}{1--15},
  \doiprefix\url{10.1007/s11051-021-05164-x} (\bibinfo{year}{2021}).

\bibitem{Yang2022}
\bibinfo{author}{Yang, X.} \emph{et~al.}
\newblock \bibinfo{journal}{\bibinfo{title}{Applications of magnetic particle
  imaging in biomedicine: Advancements and prospects}}.
\newblock {\emph{\JournalTitle{Frontiers in Physiology}}}
  \textbf{\bibinfo{volume}{13}}, \doiprefix\url{10.3389/fphys.2022.898426}
  (\bibinfo{year}{2022}).

\bibitem{Gleich}
\bibinfo{author}{Gleich, B.} \& \bibinfo{author}{Weizenecker, J.}
\newblock \bibinfo{journal}{\bibinfo{title}{Tomographic imaging using the
  nonlinear response of magnetic particles}}.
\newblock {\emph{\JournalTitle{Nature}}} \textbf{\bibinfo{volume}{435}},
  \bibinfo{pages}{1214--1217} (\bibinfo{year}{2005}).

\bibitem{Rahmer:2009uh}
\bibinfo{author}{Rahmer, J.}, \bibinfo{author}{Weizenecker, J.},
  \bibinfo{author}{Gleich, B.} \& \bibinfo{author}{Borgert, J.}
\newblock \bibinfo{journal}{\bibinfo{title}{Signal encoding in magnetic
  particle imaging: properties of the system function}}.
\newblock {\emph{\JournalTitle{BMC Medical Imaging}}}
  \textbf{\bibinfo{volume}{9}}, \bibinfo{pages}{4},
  \doiprefix\url{10.1186/1471-2342-9-4} (\bibinfo{year}{2009}).

\bibitem{BORGERT2012}
\bibinfo{author}{Borgert, J.} \emph{et~al.}
\newblock \bibinfo{journal}{\bibinfo{title}{Fundamentals and applications of
  magnetic particle imaging}}.
\newblock {\emph{\JournalTitle{Journal of Cardiovascular Computed Tomography}}}
  \textbf{\bibinfo{volume}{6}}, \bibinfo{pages}{149--153},
  \doiprefix\url{https://doi.org/10.1016/j.jcct.2012.04.007}
  (\bibinfo{year}{2012}).

\bibitem{Wells2020}
\bibinfo{author}{Wells, J.} \emph{et~al.}
\newblock \bibinfo{journal}{\bibinfo{title}{Lissajous scanning magnetic
  particle imaging as a multifunctional platform for magnetic hyperthermia
  therapy}}.
\newblock {\emph{\JournalTitle{Nanoscale}}} \textbf{\bibinfo{volume}{12}},
  \bibinfo{pages}{18342--18355}, \doiprefix\url{10.1039/D0NR00604A}
  (\bibinfo{year}{2020}).

\bibitem{Zhou2018}
\bibinfo{author}{Zhou, X.~Y.} \emph{et~al.}
\newblock \bibinfo{journal}{\bibinfo{title}{Magnetic particle imaging for
  radiation-free, sensitive and high-contrast vascular imaging and cell
  tracking.}}
\newblock {\emph{\JournalTitle{Curr Opin Chem Biol}}}
  \textbf{\bibinfo{volume}{45}}, \bibinfo{pages}{131--138},
  \doiprefix\url{10.1016/j.cbpa.2018.04.014} (\bibinfo{year}{2018}).

\bibitem{Keselman_2017}
\bibinfo{author}{Keselman, P.} \emph{et~al.}
\newblock \bibinfo{journal}{\bibinfo{title}{Tracking short-term biodistribution
  and long-term clearance of {SPIO} tracers in magnetic particle imaging}}.
\newblock {\emph{\JournalTitle{Physics in Medicine \& Biology}}}
  \textbf{\bibinfo{volume}{62}}, \bibinfo{pages}{3440--3453},
  \doiprefix\url{10.1088/1361-6560/aa5f48} (\bibinfo{year}{2017}).

\bibitem{Arami2015}
\bibinfo{author}{Arami, H.}, \bibinfo{author}{Khandhar, A.},
  \bibinfo{author}{Liggitt, D.} \& \bibinfo{author}{Krishnan, K.~M.}
\newblock \bibinfo{journal}{\bibinfo{title}{In vivo delivery, pharmacokinetics,
  biodistribution and toxicity of iron oxide nanoparticles.}}
\newblock {\emph{\JournalTitle{Chem Soc Rev}}} \textbf{\bibinfo{volume}{44}},
  \bibinfo{pages}{8576--607}, \doiprefix\url{10.1039/c5cs00541h}
  (\bibinfo{year}{2015}).

\bibitem{Nowak2022}
\bibinfo{author}{Nowak-Jary, J.} \& \bibinfo{author}{Machnicka, B.}
\newblock \bibinfo{journal}{\bibinfo{title}{Pharmacokinetics of magnetic iron
  oxide nanoparticles for medical applications}}.
\newblock {\emph{\JournalTitle{J Nanobiotechnology}}}
  \textbf{\bibinfo{volume}{20}}, \bibinfo{pages}{305--335},
  \doiprefix\url{10.1186/s12951-022-01510-w} (\bibinfo{year}{2022}).

\bibitem{Liu2021}
\bibinfo{author}{Liu, S.} \emph{et~al.}
\newblock \bibinfo{journal}{\bibinfo{title}{Long circulating tracer tailored
  for magnetic particle imaging.}}
\newblock {\emph{\JournalTitle{Nanotheranostics}}}
  \textbf{\bibinfo{volume}{5}}, \bibinfo{pages}{348--361},
  \doiprefix\url{10.7150/ntno.58548} (\bibinfo{year}{2021}).

\bibitem{Rauwerdink2010}
\bibinfo{author}{Rauwerdink, A.~M.} \& \bibinfo{author}{Weaver, J.~B.}
\newblock \bibinfo{journal}{\bibinfo{title}{Viscous effects on nanoparticle
  magnetization harmonics}}.
\newblock {\emph{\JournalTitle{Journal of Magnetism and Magnetic Materials}}}
  \textbf{\bibinfo{volume}{322}}, \bibinfo{pages}{609--613},
  \doiprefix\url{10.1016/j.jmmm.2009.10.024} (\bibinfo{year}{2010}).

\bibitem{Rauwerdink2009}
\bibinfo{author}{Rauwerdink, A.~M.}, \bibinfo{author}{Hansen, E.~W.} \&
  \bibinfo{author}{Weaver, J.~B.}
\newblock \bibinfo{journal}{\bibinfo{title}{Nanoparticle temperature estimation
  in combined ac and dc magnetic fields}}.
\newblock {\emph{\JournalTitle{Physics in Medicine \& Biology}}}
  \textbf{\bibinfo{volume}{54}}, \bibinfo{pages}{L51},
  \doiprefix\url{10.1088/0031-9155/54/19/L01} (\bibinfo{year}{2009}).

\bibitem{Rauwerdink2010oct}
\bibinfo{author}{Rauwerdink, A.~M.}, \bibinfo{author}{Giustini, A.~J.} \&
  \bibinfo{author}{Weaver, J.~B.}
\newblock \bibinfo{journal}{\bibinfo{title}{Simultaneous quantification of
  multiple magnetic nanoparticles}}.
\newblock {\emph{\JournalTitle{Nanotechnology}}} \textbf{\bibinfo{volume}{21}},
  \bibinfo{pages}{455101}, \doiprefix\url{10.1088/0957-4484/21/45/455101}
  (\bibinfo{year}{2010}).

\bibitem{BuchFranke}
\bibinfo{author}{Franke, J.} \& \bibinfo{author}{Chacon-Caldera, J.}
\newblock \bibinfo{title}{Chapter 11 - magnetic particle imaging}.
\newblock In \bibinfo{editor}{Tishin, A.~M.} (ed.)
  \emph{\bibinfo{booktitle}{Magnetic Materials and Technologies for Medical
  Applications}}, Woodhead Publishing Series in Electronic and Optical
  Materials, \bibinfo{pages}{339--393},
  \doiprefix\url{10.1016/B978-0-12-822532-5.00015-7}
  (\bibinfo{publisher}{Woodhead Publishing}, \bibinfo{year}{2022}).

\bibitem{Stehning}
\bibinfo{author}{Stehning, C.}, \bibinfo{author}{Gleich, B.} \emph{et~al.}
\newblock \bibinfo{journal}{\bibinfo{title}{Simultaneous magnetic particle
  imaging ({MPI}) and temperature mapping using multi-color mpi}}.
\newblock {\emph{\JournalTitle{International Journal on Magnetic Particle
  Imaging}}}  (\bibinfo{year}{2016}).

\bibitem{Buchholz2022}
\bibinfo{author}{Buchholz, O.} \emph{et~al.}
\newblock \bibinfo{journal}{\bibinfo{title}{{MPI}-based spatio-temporal
  estimation of a temperature profile induced by an {IR} laser}}.
\newblock {\emph{\JournalTitle{IWMPI 2022 - International Workshop on Magnetic
  Particle Imaging}}} \textbf{\bibinfo{volume}{8}}, \bibinfo{pages}{Supp 1, ID
  2203046}, \doiprefix\url{10.18416/ijmpi.2022.2203046} (\bibinfo{year}{2022}).

\bibitem{MA200433}
\bibinfo{author}{Ma, M.} \emph{et~al.}
\newblock \bibinfo{journal}{\bibinfo{title}{Size dependence of specific power
  absorption of {Fe3O4} particles in {AC} magnetic field}}.
\newblock {\emph{\JournalTitle{Journal of Magnetism and Magnetic Materials}}}
  \textbf{\bibinfo{volume}{268}}, \bibinfo{pages}{33--39},
  \doiprefix\url{10.1016/S0304-8853(03)00426-8} (\bibinfo{year}{2004}).

\bibitem{ROSENSWEIG2002}
\bibinfo{author}{Rosensweig, R.}
\newblock \bibinfo{journal}{\bibinfo{title}{Heating magnetic fluid with
  alternating magnetic field}}.
\newblock {\emph{\JournalTitle{Journal of Magnetism and Magnetic Materials}}}
  \textbf{\bibinfo{volume}{252}}, \bibinfo{pages}{370--374},
  \doiprefix\url{https://doi.org/10.1016/S0304-8853(02)00706-0}
  (\bibinfo{year}{2002}).
\newblock \bibinfo{note}{Proceedings of the 9th International Conference on
  Magnetic Fluids}.

\bibitem{Wang2005}
\bibinfo{author}{Wang, X.}, \bibinfo{author}{Gu, H.} \& \bibinfo{author}{Yang,
  Z.}
\newblock \bibinfo{journal}{\bibinfo{title}{The heating effect of magnetic
  fluids in an alternating magnetic field}}.
\newblock {\emph{\JournalTitle{Journal of Magnetism and Magnetic Materials}}}
  \textbf{\bibinfo{volume}{293}}, \bibinfo{pages}{334--340},
  \doiprefix\url{10.1016/j.jmmm.2005.02.028} (\bibinfo{year}{2005}).
\newblock \bibinfo{note}{Proceedings of the Fifth International Conference on
  Scientific and Clinical Apllications of Magnetic Carriers}.

\bibitem{guardia}
\bibinfo{author}{Guardia, P.} \emph{et~al.}
\newblock \bibinfo{journal}{\bibinfo{title}{Water-soluble iron oxide nanocubes
  with high values of specific absorption rate for cancer cell hyperthermia
  treatment}}.
\newblock {\emph{\JournalTitle{ACS Nano}}} \textbf{\bibinfo{volume}{6}},
  \bibinfo{pages}{3080--3091}, \doiprefix\url{10.1021/nn2048137}
  (\bibinfo{year}{2012}).
\newblock \bibinfo{note}{PMID: 22494015}, \eprint{10.1021/nn2048137}.

\bibitem{Jordan}
\bibinfo{author}{Jordan, A.} \emph{et~al.}
\newblock \bibinfo{journal}{\bibinfo{title}{Endocytosis of dextran and
  silan-coated magnetite nanoparticles and the effect of intracellular
  hyperthermia on human mammary carcinoma cells in vitro}}.
\newblock {\emph{\JournalTitle{Journal of Magnetism and Magnetic Materials}}}
  \textbf{\bibinfo{volume}{194}}, \bibinfo{pages}{185--196},
  \doiprefix\url{10.1016/S0304-8853(98)00558-7} (\bibinfo{year}{1999}).

\bibitem{Liu}
\bibinfo{author}{Liu, X.~L.} \emph{et~al.}
\newblock \bibinfo{journal}{\bibinfo{title}{Optimization of surface coating on
  {Fe3O4} nanoparticles for high performance magnetic hyperthermia agents}}.
\newblock {\emph{\JournalTitle{J. Mater. Chem.}}}
  \textbf{\bibinfo{volume}{22}}, \bibinfo{pages}{8235--8244},
  \doiprefix\url{10.1039/C2JM30472D} (\bibinfo{year}{2012}).

\bibitem{Tong}
\bibinfo{author}{Tong, S.}, \bibinfo{author}{Quinto, C.~A.},
  \bibinfo{author}{Zhang, L.}, \bibinfo{author}{Mohindra, P.} \&
  \bibinfo{author}{Bao, G.}
\newblock \bibinfo{journal}{\bibinfo{title}{Size-dependent heating of magnetic
  iron oxide nanoparticles}}.
\newblock {\emph{\JournalTitle{ACS Nano}}} \textbf{\bibinfo{volume}{11}},
  \bibinfo{pages}{6808--6816}, \doiprefix\url{10.1021/acsnano.7b01762}
  (\bibinfo{year}{2017}).
\newblock \bibinfo{note}{PMID: 28625045}.

\bibitem{Prashant}
\bibinfo{author}{Chandrasekharan, P.} \emph{et~al.}
\newblock \bibinfo{journal}{\bibinfo{title}{Vitamin e
  (d-alpha-tocopheryl-co-poly(ethylene glycol) 1000 succinate)
  micelles-superparamagnetic iron oxide nanoparticles for enhanced
  thermotherapy and {MRI}}}.
\newblock {\emph{\JournalTitle{Biomaterials}}} \textbf{\bibinfo{volume}{32}},
  \bibinfo{pages}{5663--5672},
  \doiprefix\url{10.1016/j.biomaterials.2011.04.037} (\bibinfo{year}{2011}).

\bibitem{Tay}
\bibinfo{author}{Tay, Z.~W.} \emph{et~al.}
\newblock \bibinfo{journal}{\bibinfo{title}{Magnetic particle imaging-guided
  heating in vivo using gradient fields for arbitrary localization of magnetic
  hyperthermia therapy}}.
\newblock {\emph{\JournalTitle{ACS Nano}}} \textbf{\bibinfo{volume}{12}},
  \bibinfo{pages}{3699--3713}, \doiprefix\url{10.1021/acsnano.8b00893}
  (\bibinfo{year}{2018}).
\newblock \bibinfo{note}{PMID: 29570277}, \eprint{10.1021/acsnano.8b00893}.

\bibitem{wei2020}
\bibinfo{author}{Wei, H.}, \bibinfo{author}{Behrends, A.},
  \bibinfo{author}{Friedrich, T.} \& \bibinfo{author}{Buzug, T.}
\newblock \bibinfo{journal}{\bibinfo{title}{A heating coil insert for a
  preclinical mpi scanner}}.
\newblock {\emph{\JournalTitle{International Workshop on Magnetic Particle
  Imaging (IWMPI) 2017, Book of Abstracts}}} \textbf{\bibinfo{volume}{6}},
  \bibinfo{pages}{1--3}, \doiprefix\url{10.18416/IJMPI.2020.2009056}
  (\bibinfo{year}{2020}).

\bibitem{Kuboyabu}
\bibinfo{author}{Kuboyabu, T.} \emph{et~al.}
\newblock \bibinfo{journal}{\bibinfo{title}{Magnetic particle imaging for
  magnetic hyperthermia treatment: Visualization and quantification of the
  intratumoral distribution and temporal change of magnetic nanoparticles}}.
\newblock {\emph{\JournalTitle{Open Journal of Medical Imaging}}}
  \textbf{\bibinfo{volume}{6}}, \bibinfo{pages}{1--15},
  \doiprefix\url{10.4236/ojmi.2016.61001} (\bibinfo{year}{2016}).

\bibitem{textbook}
\bibinfo{author}{Marieb, E.} \& \bibinfo{author}{Hoehn, K.}
\newblock \emph{\bibinfo{title}{Human Anatomy \& Physiology, eBook, Global
  Edition}} (\bibinfo{publisher}{Pearson Education}, \bibinfo{year}{2015}).

\bibitem{Murase2015}
\bibinfo{author}{Murase, K.} \emph{et~al.}
\newblock \bibinfo{journal}{\bibinfo{title}{Usefulness of magnetic particle
  imaging for predicting the therapeutic effect of magnetic hyperthermia}}.
\newblock {\emph{\JournalTitle{Open Journal of Medical Imaging}}}
  \textbf{\bibinfo{volume}{05}}, \bibinfo{pages}{85--99},
  \doiprefix\url{10.4236/ojmi.2015.52013} (\bibinfo{year}{2015}).

\bibitem{Malhotra2019}
\bibinfo{author}{Malhotra, A.} \emph{et~al.}
\newblock \bibinfo{journal}{\bibinfo{title}{Tracking the growth of
  superparamagnetic nanoparticles with an in-situ magnetic particle
  spectrometer (inspect)}}.
\newblock {\emph{\JournalTitle{Scientific Reports}}}
  \textbf{\bibinfo{volume}{9}}, \bibinfo{pages}{10538},
  \doiprefix\url{10.1038/s41598-019-46882-6} (\bibinfo{year}{2019}).

\bibitem{behrends2019selfcompensating}
\bibinfo{author}{Behrends, A.}, \bibinfo{author}{Wei, H.},
  \bibinfo{author}{Friedrich, T.}, \bibinfo{author}{Neumann, A.} \&
  \bibinfo{author}{Buzug, T.~M.}
\newblock \bibinfo{title}{A self-compensating coil setup for combined magnetic
  particle imaging and magnetic fluid hyperthermia}.
\newblock In \emph{\bibinfo{booktitle}{9th International Workshop on Magnetic
  Particle Imaging}} (\bibinfo{year}{2019}).

\bibitem{Chen:2004vj}
\bibinfo{author}{Chen, Z.-J.} \emph{et~al.}
\newblock \bibinfo{journal}{\bibinfo{title}{A realistic brain tissue phantom
  for intraparenchymal infusion studies.}}
\newblock {\emph{\JournalTitle{J Neurosurg}}} \textbf{\bibinfo{volume}{101}},
  \bibinfo{pages}{314--322}, \doiprefix\url{10.3171/jns.2004.101.2.0314}
  (\bibinfo{year}{2004}).

\end{thebibliography}

\section*{Acknowledgements}
This work was supported by the German Federal Ministry of Research project FMT 13GW0230. The authors thank the research and core facility AMIR of the Medical Faculty, University of Freiburg, for support in MRI imaging.

\section*{Author contributions statement}
OB, KS, JF, CM, UGH and SB conducted and conceived experiments. HW designed and constructed the hyperthermia insert. UGH and SB supervised the study. UGH directed the project FMT. CG synthesized SPION samples specifically for the study. UGH, TB, PL and JF secured funding for this project. All authors contributed to discussing, writing and revising the manuscript and approved the final version.

\section*{Additional information}

\textbf{Accession codes} Data and programing code will be provided upon reasonable request. \\
\textbf{Competing interests}
KS and JF are employed by Bruker BioSpin MRI GmbH. CG is employed by micromod Partikeltechnologie GmbH. All other authors have no competing interests to report.

\begin{figure}[ht]
\centering
\includegraphics[width=\linewidth]{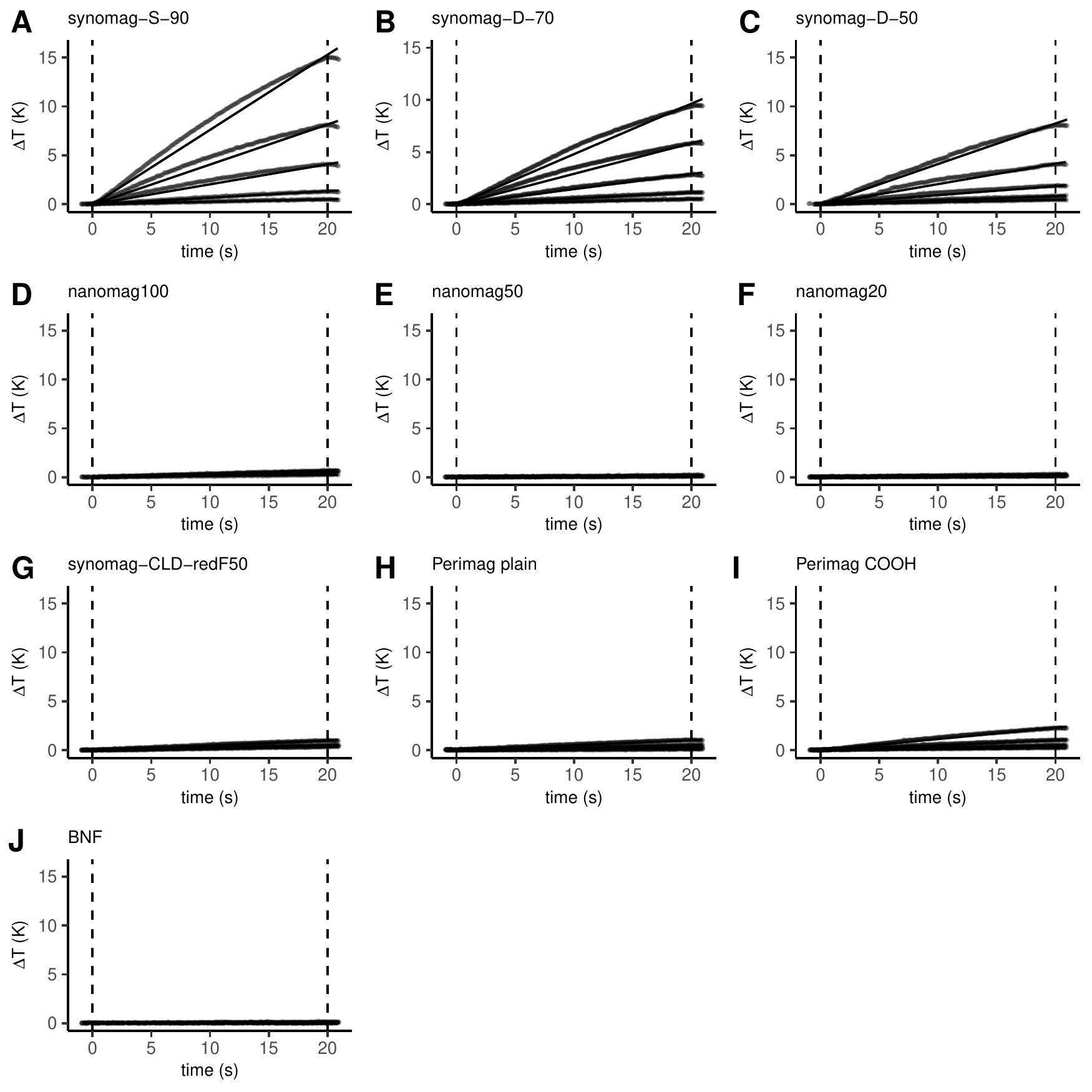}
\caption{Supplementary material: Heating slopes of different SPIONs at different iron concentrations during MFH application}
\label{fig:tracer slopes}
\end{figure}

\begin{figure}[ht]
\centering
\includegraphics[width=\linewidth]{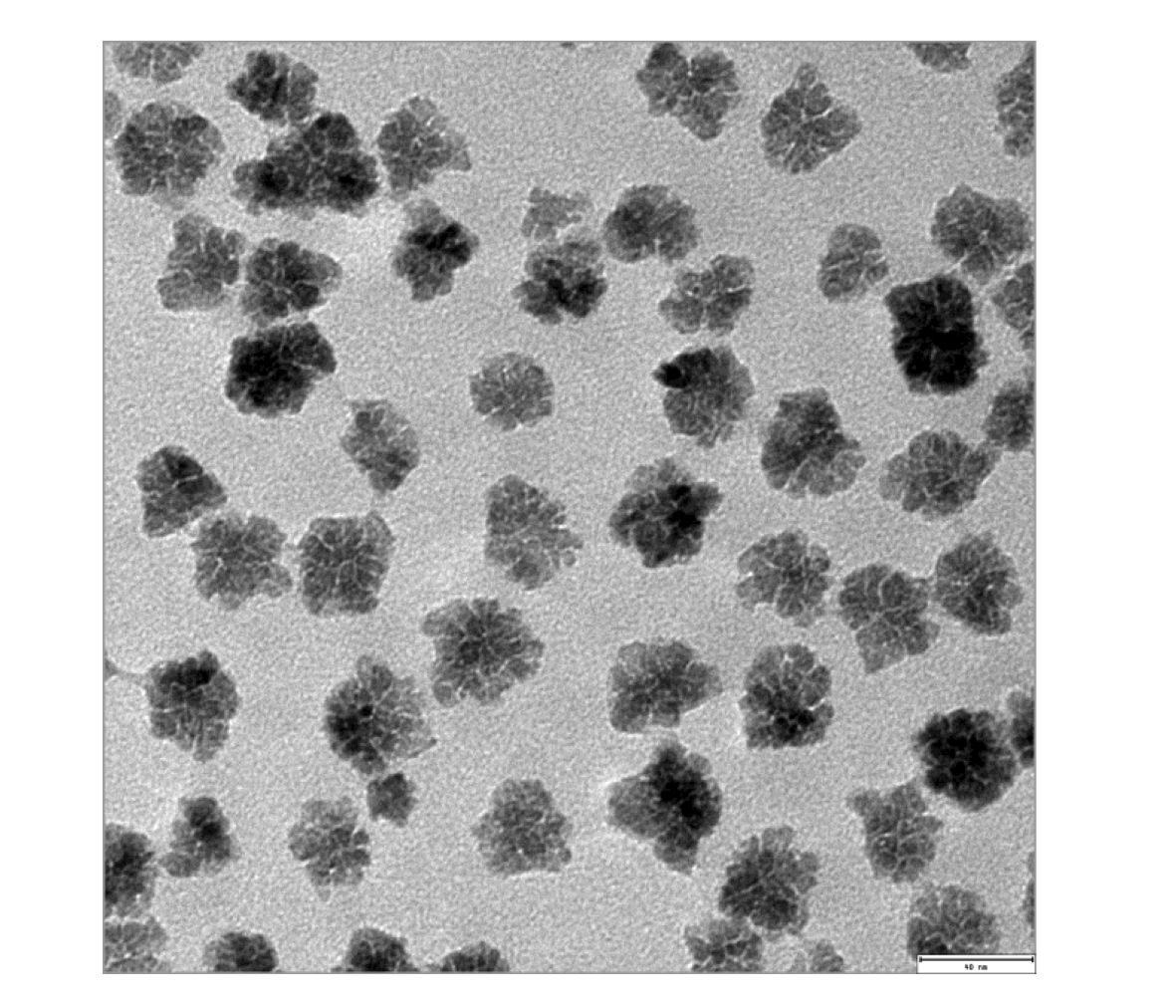}
\caption{Supplementary material2: examplary TEM image of synomag-S-90 SPION}
\label{fig:tracer TEM}
\end{figure}

\end{document}